\documentclass[11pt]{article}
\usepackage{epsfig,a4wide}
\def\vep{\varepsilon}

\def\mev{\,{\rm MeV}}
\def\gev{\,{\rm GeV}}

\renewcommand{\c}{\rm{c}}
\newcommand{\g}{\rm{g}}
\newcommand{\s}{\rm{s}}
\newcommand{\q}{\rm{q}}
\newcommand{\cbar}{\overline{\rm{c}}}

\def\cbc{\rm{c}\overline{\rm c}}
\def\qbq{{\rm q}\overline{\rm q}}
\def\ubu{{\rm u}\overline{\rm u}}
\def\dbd{{\rm d}\overline{\rm d}}
\def\sbs{{\rm s}\overline{\rm s}}

\def\BbB{B\overline{B}}
\def\PbP{P P }
\def\VbV{V V }
\def\pbp{p\overline{p}}
\def\jp{J/\psi}
\def\jpomphi{\jp-\omega-\phi}
\def\mjp{M_{J/\psi}}
\newcommand{\mc}{m_{\c}}
\def\als{\alpha_s}
\def\su3{{\rm SU}(3)_{{\rm F}}}

\begin{document}

%%%%%%%%%%%%%%%%%%%%%%%%%%%%

%%%%% Extra Title Page %%%%%%%%%%%%%%

%\sloppy
%\renewcommand{\arraystretch}{1.5}
\thispagestyle{empty}
{%\large
\begin{flushright}
WU-B 00--01 \\
PITHA 00/05 \\
March 2000\\
hep-ph/0003096
\end{flushright}
\vspace{\fill}
\begin{center}
{\LARGE\bf Implications of light-quark admixtures on \\ 
       charmonium decays into meson pairs }\\[10mm]
{\Large Thorsten Feldmann$^1$ and Peter Kroll$^2$} \\[3mm]
{\small\it ${}^1$Institut f\"ur Theoretische Physik~E, RWTH Aachen,
  52056 Aachen, Germany}
\\[1mm]
{\small\it ${}^2$Fachbereich Physik, Universit\"at Wuppertal,
42097 Wuppertal, Germany}\\[1mm]
%{ E-mail: kroll@theorie.physik.uni-wuppertal.de}\\[2ex]
\end{center}
\vspace{\fill}

\begin{center}
{\bf Abstract}\\[2ex]
\begin{minipage}{13cm}
We argue that charmonium decays into meson pairs fall into two
distinct classes: one that is under control of perturbative QCD
and another one that is governed by a soft mechanism. We
concentrate on a systematic analysis of $\jp (\Psi')$ decays
into a light pseudoscalar and a light vector meson and 
$\eta_c$ decays into a pair of light vector mesons.
These processes belong to the second class and are
characterized by non-conserved hadronic helicity. It is assumed that,
in these cases, the charmonium state decays dominantly through a light-quark
Fock component by a soft mechanism which is characteristic of OZI-rule
allowed strong decays. Estimating the light-quark admixture by meson
mixing, we obtain a reasonable description of the branching ratios for
these processes.  
\end{minipage}
\end{center}

\vskip5em
\begin{center}
(submitted to Physical Review D)
\end{center}

\vspace{\fill}
}
\clearpage

%%%%%%%%%%%%%%%%%%%%%%%%%%%%%%%%%%%%%%%%%%%%%%%%%%%%%%%%%%%%%%%%%%%%
\section{Introduction}
%%%%%%%%%%%%%%%%%%%%%%%%%%%%%%%%%%%%%%%%%%%%%%%%%%%%%%%%%%%%%%%%%%%
\label{sec:intro}

Exclusive charmonium decays have been investigated within perturbative
QCD by many authors, e.g.\ \cite{dun80,BrL81,che82}. It has been argued
that the dominant dynamical mechanism is $\c\cbar$ annihilation into the
minimum number of gluons allowed by color conservation and charge
conjugation, and subsequent creation of light quark-antiquark pairs
forming the final state hadrons. 
This factorization of long- and short-distance physics has been shown to
hold in the formal limit $\mc\to\infty$.  
The dominance of annihilation through gluons is most strikingly reflected
in the narrow widths of charmonium decays into hadronic channels in a
mass region where strong decays typically have widths of hundreds of
MeV \cite{app75}. Since the $\c$ and the $\cbar$ quarks only
annihilate if their mutual distance is less than about $1/\mc$ (where
$\mc$ is the $\c$-quark mass) which is smaller than the
non-perturbative charmonium radius, and since the average virtuality of the
gluons is of the order of $1 - 2\, \gev^2$ one may indeed expect
perturbative QCD to be at work although corrections are presumably
substantial. The charm-quark mass is too small in order to suppress
power corrections decisively although it is large enough to allow
perturbative QCD (pQCD) calculations. The bottomonium system, for which no
exclusive hadronic decay has been observed as yet, should exhibit the
pattern of perturbative predictions much cleaner.
 
In hard exclusive reactions higher Fock state contributions are
usually suppressed by inverse powers of the hard scale, $Q$, appearing
in the process ($Q=\mc$ for exclusive charmonium decays), as compared
to the valence Fock state contribution. Hence, higher Fock state 
contributions are expected to be negligible in most cases. For exclusive 
charmonium decays, however, the 
valence Fock state contributions are often suppressed for one or the other
reason. In such a case higher Fock state contributions or other
peculiar contributions such as power corrections or small components
of the hadronic wave functions may become important. 
One such exception are the exclusive decays of the $\chi_{\c J}$
mesons for which the $\c\cbar$ pair forms a color-singlet ${\rm P}$-wave
state in the valence Fock state (notation: $\c\cbar_{1}(^3P_J)$). As
has been shown in Ref.~\cite{BKS1} the next-higher Fock state,
$\c\cbar \g$, where the quark-antiquark pair forms a
$\c\cbar_{8} ({}^3{\rm S} _1)$ state, is not suppressed as compared to the
valence contribution. Therefore, the 
neglect of the $\c\cbar \g$ contributions, which are customarily referred
to as the color-octet contributions \cite{BBL}, is unjustified in the $\chi_{\c J}$
decays. Indeed as has been discussed in Ref.~\cite{BKS1} for the
$\PbP$ channels (where $P$ denotes a pseudoscalar meson) and in 
Ref.~\cite{wong} for their decays into baryon-antibaryon pairs the
color-octet contributions are substantial and are definitely needed
for obtaining agreement with experiment. 

Decays into final states involving vector mesons ($V$) have often
peculiar properties too. On general grounds,
only hadronic helicity non-conserving amplitudes contribute to some of
these processes. Hence, these reactions are not under control of
leading-twist pQCD. A famous example of the helicity non-conserving
processes is set by the $\jp$ and $\Psi'$ decays into $\rho\pi$. While
$\jp\to\rho\pi$ has the largest branching ratio of all two-body
hadronic $\jp$ decays, the branching ratio of $\Psi'\to\rho\pi$ is
very small \cite{PDG}. The combination of these two experimental facts
forms the so-called $\rho\pi$ puzzle. Many of the other $PV$ channels
behave similarly. There
are many attempts to understand the $\jp (\Psi') \to PV$ decays;
the proposed mechanisms reach {}from intrinsic charm in the light mesons
\cite{bro97}, to color-octet contributions \cite{chen98}, 
vector meson mixing \cite{cla83,pin90}, final
state interactions \cite{li97,suz} and admixtures of a glueball nearly
degenerate with the $\jp$ \cite{hou83,tua99}. Another example of a
helicity non-conserving decay is set by the $\eta_{\c}$ decay into
proton and antiproton. In Ref.~\cite{kro:93a} it is attempted to
explain this process through diquarks as quasi-elementary constituents of 
baryons. 

In this article we are going to attempt a systematic study of the hadronic
helicity non-conserving $\jp (\Psi') \to PV$ decays. We will assume
that, with a small probability, the charmonium possesses Fock
components built from light quarks only. Through these Fock components
the charmonium state decays by a soft mechanism. 
We will model this decay mechanism by $\jpomphi$ mixing and
subsequent $\omega$ (or $\phi$) decay into the $PV$ state. This, in
the absence of the leading-twist perturbative QCD contribution,
dominant mechanism is to be supplemented by the electromagnetic decay
contribution and, depending on quantum numbers, by an anomalous,
doubly Okubo-Zweig-Iizuka- (OZI-) rule violating contribution. 
We are not able to calculate
the amplitudes of these decay processes directly but, employing
flavor symmetry and mixing schemes for vector and pseudoscalar
mesons, we find relations among them. With a few parameters, adjusted 
to experiment, we thus obtain a scheme of classification which
comprises an understanding of these decay processes. The analysis of 
the $\jp (\Psi')\to PV$ processes along these lines is partly to be
considered as an update of previous work
\cite{chen98,tua99,sei88,bra97} where a general parametrization of the
amplitudes for $\jp (\Psi')\to PV$ is fitted to experiment. We
improve this parametrization by including new ideas on meson mixing
\cite{FKS1} and by considering recent experimental results {}from the BES
collaboration \cite{BES}. We also present a, partially even
quantitative, physical interpretation of our parametrization which
differs {}from those presented elsewhere \cite{chen98,tua99,sei88}.
Also new in our work is the extension of the
mixing approach to the $\eta_{\c} \to \VbV$ decays. 

The paper is organized as follows: Qualitative features of charmonium
decays into pairs of mesons are discussed in Sec.~\ref{sec:2}. Next,
in 
Sec.~\ref{sec:3}, 
we investigate the $\jp$ and $\Psi'$ decays into the $PV$ channels
by applying meson mixing. The mixing mechanism is also used to analyze
the $\eta_{\c}\to \VbV$ decays (Sec.~\ref{sec:4}). 
In Sec.~\ref{sec:5} we will comment on
the mixing contributions to ${\rm S}$-wave charmonium  decays into
baryons and antibaryons briefly before we present our concluding remarks
in Sec.~\ref{sec:sum}. Details of our treatment of vector meson mixing are given
in the Appendix. 

%%%%%%%%%%%%%%%%%%%%%%%%%%%%%%%%%%%%%%%%%%%%%%%%%%%%%%%%%%%%%%%%%%%%%
\section{Qualitative features of charmonium decays into meson pairs}
\label{sec:2}
%%%%%%%%%%%%%%%%%%%%%%%%%%%%%%%%%%%%%%%%%%%%%%%%%%%%%%%%%%%%%%%%%%%%

Let us consider the $\PbP$, $PV$
and $\VbV$ final states. A few of these channels are strictly forbidden
by angular momentum and parity conservation, see Table~\ref{tab:dec}. 
Several other channels, characterized by non-conserved
naturalness,\footnote{
The naturalness of a meson $i$ having spin $J_i$ and parity $P_i$
 is defined as
$\sigma_i=(-1)^{J_i}\,P_i$ \cite{CZ84}.}
$\sigma_{\c} \neq \sigma_1 \sigma_2$, are forbidden in pQCD to
leading-twist 
order; i.e.\ higher-twist or
other dynamical mechanisms are at work here. This
comes about for the following reasons: The helicity amplitudes for
these processes are proportional to the totally antisymmetric
$\epsilon$-tensor contracted, in all possible ways, with the available
Lorentz vectors, namely the two independent momenta, $p_1$ and $p_2$,
the polarization vector(s) of the light vector meson(s),
$\varepsilon_i$, and, with the exception of the $\eta_{\c}$, the
polarization vector or tensor\footnote{ 
Since the polarization tensor is symmetric only one Lorentz index can be
contracted with the $\epsilon$-tensor.}
of the charmonium state. Now, in the rest frame of the decaying meson,
the polarization vector of a helicity zero vector (or axial vector)
meson can be expressed as a linear combination of the two final state
momenta regardless whether or not the mass of the light vector meson
is neglected. Hence, the number of independent Lorentz vectors is
insufficient to contract the $\epsilon$-tensor with the consequence of
vanishing amplitudes for processes involving longitudinally polarized
vector mesons. Moreover, $\varepsilon_1^*(\pm1)=-\varepsilon_2^*(\mp1)$
holds in the rest frame of the decaying meson. As a consequence of these
properties the only non-zero helicity amplitudes for the processes
with non-conserved naturalness are those where $\lambda_1+\lambda_2\neq
0$. In other words vector mesons are transversely polarized in
reactions with non-conserved naturalness.\footnote{ Somewhat exceptional is the
$\chi_{\c 1}\to \VbV$ case where one out of the three spin-1 mesons
may be longitudinally polarized.} We thus conclude that in these
processes hadronic helicity conservation
\begin{equation}
   \lambda_1 + \lambda_2 = 0
\label{hsr}
\end{equation}
is violated.
Helicity conservation is not a consequence of a particular symmetry
but is a dynamical consequence of leading-twist perturbative QCD 
(i.e.\ using leading-twist wave functions and valence Fock states only):
The virtual gluons {}from the annihilation of the $\c\cbar$ pair create 
the light, (almost) massless quarks and antiquarks in opposite
helicity states. To the extent that the hadronic wave functions do not
embody any non-zero orbital angular momentum components, the quark 
helicities sum up to their parent hadron's helicity. Hence, the total 
helicity of the final state hadrons is zero. Consequently, 
processes that violate helicity conservation, are not governed by
leading-twist pQCD. The remaining two-meson 
channels, marked by ticks in Table~\ref{tab:dec}, are accessible to a
perturbative treatment. 
%%%%%%%%%%%%%%%%%%%%%%%%%%%%%%%%%%%%%%%%%%%%%%%%%%%%%%%%%%%%%%%%%%%%%
% TABLE 1
%%%%%%%%%%%%%%%%%%%%%%%%%%%%%%%%%%%%%%%%%%%%%%%%%%%%%%%%%%%%%%%%%%%%%
\begin{table}
\begin{center}
  \begin{tabular}{|c||c|c|c|} \hline 
              &  $\;\PbP\;$&  $\;PV\;$    &  $\;\VbV\;$    \\ \hline \hline 
   $\eta_{\c}$&   -    &$(\surd)$  &$\epsilon$  \\ \hline 
     $\jp$    &($\surd$)&$\epsilon$&($\surd$)    \\ \hline 
 $\chi_{\c 0}$&$\surd$&    -     &$\surd$    \\ \hline 
 $\chi_{\c 1}$&   -    &($\surd)$  &$\epsilon$  \\ \hline 
 $\chi_{\c 2}$&$\surd$&$(\epsilon)$ &$\surd$    \\ \hline 
% $h_{\c}$     &   -    & $\surd$  & $\epsilon$ \\ \hline 
   \end{tabular}
\end{center}
 \caption[]{Charmonium decays into $\PbP$, $PV$ and $\VbV$ meson pairs.
            The symbols ($-$, $\epsilon$, $\surd$) denote by angular
            momentum and parity conservation forbidden, in pQCD to
            leading-twist order forbidden and allowed channels,
            respectively. The brackets indicate that these channels
            violate either $G$-parity or isospin invariance for
            non-strange mesons.}
\label{tab:dec}
\end{table}
%%%%%%%%%%%%%%%%%%%%%%%%%%%%%%%%%%%%%%%%%%%%%%%%%%%%%%%%%%%%%%%%%%%%

\begin{figure}[bt]
\begin{center}
\psfig{file=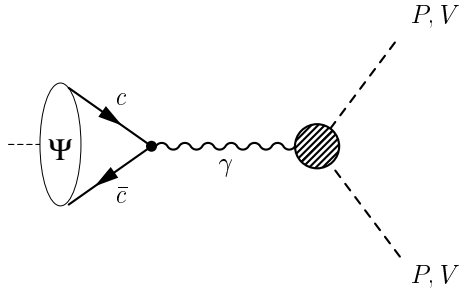, bb=60 640 260 775, height=4cm}
\end{center}
\caption{Electromagnetic $\jp$ and $\Psi'$ decays into meson
pairs. The shaded circle indicates the time-like electromagnetic
(transition) form factor of light vector and/or pseudoscalar mesons.}
\label{fig:elm} 
\end{figure}

Next let us consider $G$-parity and isospin. $G$-parity or 
isospin-violating decays are not strictly forbidden since they can proceed 
through electromagnetic $\c\cbar$ annihilation and may receive
contributions {}from the isospin-violating part of QCD. The latter 
contributions, being of the order of quark mass differences, seem to 
be small \cite{che82}. $G$-parity or isospin-violating decays of
$C$-even charmonia (e.g.\ $\eta_{\c},\chi_{\c J} \to PV$ ($J=1,2$) 
for non-strange final state mesons) have not been observed
experimentally as yet \cite{PDG}. Proceeding on the assumption that
these decays are dominantly mediated by $\c\cbar\to 2\gamma^*\to
PV$, this is understandable. They 
are then suppressed by a factor $(\alpha_{\rm em}/\als)^4$ as compared 
to the $G$-parity and isospin allowed decays of the $C$-even charmonia 
and their decay widths are therefore extremely small. Channels 
involving strange mesons (e.g.\ $K K^*$), should also be strongly suppressed by 
virtue of $U$-spin invariance. For $\jp$
decays the situation is different. Many $G$-parity violating (e.g.\
$\pi^+\pi^-$) or isospin-violating (e.g.\ $\omega\pi$) decays have
been observed, the experimental branching ratios being of the order 
of $10^{-4} - 10^{-3}$ \cite{PDG}. As compared to $G$-parity and 
isospin allowed $\jp$ decays they are typically suppressed by factors 
of about $10^{-2} - 10^{-1}$ in accordance with what is expected for an 
electromagnetic decay mechanism\footnote{
A possible correction due to $\cbc$ annihilation mediated by
$\gamma^*g^*g^*$ is ignored by us here. \label{foot4}} 
(see Fig.~\ref{fig:elm}).

That at least two distinct dynamical mechanisms are at work in
exclusive charmonium decays can be realized {}from a comparison of $\jp$
and $\Psi'$ decays. Suppose these decays are under control of a
mechanism that respects QCD factorization and where the charm and
anticharm quarks annihilate into gluons and/or photons at small
mutual distance. The charmonium wave function{} is then 
probed at small spatial separations and, therefore, is well represented
by the corresponding decay constant, $f_{\jp}$ or $f_{\Psi'}$, which is
measured in electronic decays of that charmonium state. If this short
distance mechanism is responsible for the $\jp$ and $\Psi'$ decay into
a particular two hadron channel, the ratio
\begin{equation}
\kappa_{12}\,=\, \frac{{\cal B}(\Psi'\to h_1h_2)} {{\cal B}(\jp\to h_1h_2)}\,
            \frac{{\cal B}(\jp\to e^+e^-)} {{\cal B}(\Psi'\to e^+e^-)}\,
            \left(\frac{\varrho_{12}(\jp)}{\varrho_{12}(\Psi')}\right)^n
\label{kappa}
\end{equation}
should be close to unity\footnote{
   In a consistent perturbative calculation to lowest order in the
   $\c$-quark velocity expansion \cite{BKS1,BBL,BKS2} one has to evaluate the
   hard scattering amplitude {}from $2\mc$ instead {}from the bound state
   mass; the mass difference is an effect of order $v^2$, i.e.\
   of the same order as the
   contributions {}from the next higher Fock states. An
   additional factor of $(M_{\Psi'}/\mjp)^{2n}$ in Eq.~(\ref{kappa})
   ($n=4$ for baryon-antibaryon channels for instance), to be
   found in the literature occasionally, is inconsistent 
   in this respect.}
($n=3$ for $PV$  channels and $n=1$ otherwise). Since 
${\cal B}(\Psi'\to e^+e^-)/{\cal B}(\jp\to e^+e^-)$ amounts to about
$14\%$, the relation~(\ref{kappa}) is occasionally termed the $14\%$ rule. 
The phase space factor in Eq.~(\ref{kappa}) is defined by
\begin{equation}
\varrho_{12} (j)=\sqrt{1 - 2(m_1^2+m_2^2)/M^2_j + (m_1^2-m_2^2)^2/M_j^4}
\label{phase}
\end{equation}
where  $M_j$ is the mass of the charmonium state. It is of
numerical relevance only for particles with masses above $1
~\gev$. Experimental results for $\kappa$ are listed in
Table~\ref{tab:kappa}; 
the data are taken {}from 
Refs.~\cite{PDG,BES,bai99a,bai98c,Bai:1998fp,che99}. 
For comparison we also include
experimental results for 
the baryon-antibaryon, vector--tensor and axial-vector--pseudoscalar
channels in the table. We observe that, within
often large errors or only within bounds, $\kappa$ is indeed compatible 
with unity for the baryon-antibaryon channels, for $b_1\pi$, $K^{*0} K^{*0}$ and
for some of the $PV$ channels, notably for the $G$-parity and isospin-violating
channels $\pi\omega$, $\eta\rho$, $\eta'\rho$. For most of the
other $PV$ channels and, perhaps, to a lesser extent also for the vector-tensor
channels,
$\kappa$ is well below unity.
%\footnote{
We stress that 
the decays into a vector and a tensor meson are not forbidden by hadronic
helicity conservation,
but leading-twist pQCD feeds the helicity amplitude ${\cal
   M}_{00\lambda_c}$ only. There are many other, in general
   non-zero amplitudes that do not respect hadronic helicity
   conservation. Note that the branching ratios of $\jp$ decays into $VT$
   are large, comparable with those for the $PV$ channels while those for the
   corresponding $\Psi'$ decays are small. 
For the $\PbP$ channels the situation is
unclear; better data are demanded. We do not include the data for 
multi-particle channels in the table but we note that, with the exception of
the final states consisting of three pseudoscalars, $\kappa$ is
compatible with unity for all these channels. This fact is presumably a
consequence of an underlying perturbatively generated three gluon jet structure
which fragments into multi-particle final states. This interpretation
is supported by estimates of the total width for $\jp$ decays into
light hadrons through the decay into three real gluons \cite{nov78}. In
the case of the $\Upsilon$ decays the three jet structure is
experimentally established \cite{mey79}.

\begin{figure}[bt]
\begin{center}
\psfig{file=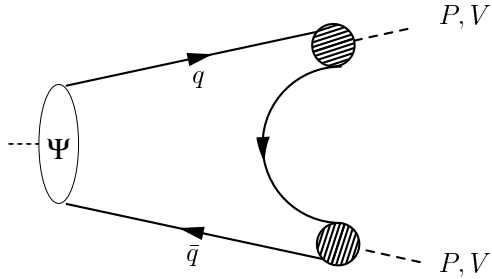, bb=60 640 270 770, height=4cm }
\end{center}
\caption{Schematic picture of the mixing mechanism for charmonium
decays into meson pairs.}
\label{fig:strong} 
\end{figure}

The small value of $\kappa$ for those 
$PV$ channels that are dominated by strong interactions, arises {}from
large $\jp$ branching ratios as an inspection of Table~\ref{tab:kappa}
reveals. The $\Psi'$ decays into $PV$, on the other hand, seem to
behave, at least in tendency, according to expectations based on pQCD;
their branching ratios are smaller than those for the leading-twist
allowed channels. Obviously, another dynamical mechanism is called for
which is active for the $\jp$ decays but suppressed for the $\Psi'$
ones. For this mechanism we assume that the charmonium state possesses  
Fock components built {}from light quarks only. It can then decay
through these Fock components by a soft mechanism that is characteristic of
OZI-rule allowed strong decays (see Fig.~\ref{fig:strong}). This mechanism is
obviously more peripheral than the short distance $\cbc$ annihilation;
i.e.\ it probes the charmonium wave function at all
quark-antiquark separations and feels therefore the difference between
a $1{\rm S}$ and a $2{\rm S}$ radial wave function. The node in the latter is
supposed to lead to a strong suppression of this mechanism in the
$\Psi'$ decays.\footnote{This assumption bears a resemblance to the node effect discussed in
   Ref.~\cite{pei98}.}
We model this mechanism by mixing of vector mesons, $\jpomphi$.
For the mixing angle, $\tilde{\theta}_{\c}$, being of order 
$1/\mc^2$, we take the value $-0.1^\circ$ which corresponds to a 
light-quark admixture to the charmonium state of order of $10^{-3}$, see
the Appendix. The light vector mesons, $\omega$ and $\phi$, couple to the
final state mesons through vertex functions, e.g.\ $g_{\omega V
P}$. In terms of vector meson mixing the suppression of the $\Psi'\to
PV$ decays follows from substantially weaker couplings
(arguments in favor of that assumption are given in 
Refs.~\cite{pin90,kra84}) of the radially excited $\omega'$ and $\phi'$
mesons (assumed to mix with the $\Psi'$)
to the ground state pseudoscalar and vector mesons and, perhaps, from 
a smaller mixing angle. 
The mixing mechanism may also be at work in the vector-tensor channels
(with vertex functions $g_{VVT}$) and may
generate the small value of $\kappa$ there, see Table~\ref{tab:kappa}.

Conversely one may also consider the $\cbc$ Fock components of 
$\omega$ and $\phi$ mesons in the final state;
their probablilities are the same as the ones for light quark
components in the $\jp$ meson,
see, Eq.~(\ref{mixprops}) in the appendix. 
The initial $\cbc$ pair from the $\jp$ may then feed that
 Fock component of the $\omega$ or $\phi$. However, the second final state
 meson is to be generated by an OZI-rule {\em violating} mechanism
 now, and is to be considered as a higher-order correction to the
leading mixing mechanism we propose.
 For channels like $\omega\eta'$ one may also think of an excitation of the
 $\cbc$ components of both the light mesons. 
 Obviously, the probability of finding a $\cbc$ pair
 in a light meson enters {\em quadratically} now, and
 this decay mechanism is suppressed by an additional power of $1/m_c^2$.
 Hence, we ignore both the above possibilities. 
 The initial $\cbc$ pair may also excite a $\cbc \qbq$
 Fock component in the light (vector) meson,
 as suggested by Brodsky and Karliner~\cite{bro97}. In that picture
 the $\qbq$ pair from this Fock component forms the other 
 final state meson. Of course, one expects the probability of this
 higher Fock state to be very small. In any case the approach of
 Brodsky and Karliner should be viewed as complementary to ours, since
 the mechanism that they propose cannot be represented by
 meson-mixing supplied with OZI-rule allowed decay vertices.
  
We note that in Refs.~\cite{chen98,tua99,sei88} the strong
interaction mechanism for the hadronic helicity non-conserving 
decays although parametrized in a
similar way as we do, is interpreted differently. Thus, Chen and
Braaten \cite{chen98} argue that this contribution is generated by an
additional Fock state gluon (the $\cbc$ pair is therefore in a
color-octet state) which carries a unit of helicity {}from the $\jp$
to one of the light mesons without participating in the hard process.
Chen and Braaten provide arguments that
the suppression of the $\Psi'$ decays
into $PV$ mesons as well as into the vector-tensor channels is due to
the energy gap between the mass of the $\Psi'$ and the 
$D\overline{D}$ threshold. 
In Refs.~\cite{tua99,sei88}, on the other hand, mixing of the $\jp$
with a fairly narrow $J^{PC} = 1^{--}$ glueball, nearly degenerate
with the $\jp$, is assumed. The $\jp$ can then decay through the glueball
while this mechanism is not available for the $\Psi'$.
Searches for such a glueball, performed by
the BES collaboration \cite{bai96}, however turned out negative
although not fully conclusive to claim the demise of the glueball
interpretation. For the final state interaction
mechanism advocated in Refs.~\cite{li97,suz} the suppression of the
$\Psi'\to PV$ decays is due to a fortuitous cancellation of various
contributions and, probably, does not hold for the vector-tensor channels. 

%%%%%%%%%%%%%%%%%%%%%%%%%%%%%%%%%%%%%%%%%%%%%%%%%%%%%%%%%%%%%%%%%%%%%
% TABLE 2
%%%%%%%%%%%%%%%%%%%%%%%%%%%%%%%%%%%%%%%%%%%%%%%%%%%%%%%%%%%%%%%%%%%%%
\begin{table}
\begin{center}
\begin{tabular}{|c||c|c||c|} \hline
             channel & $10^4 {\cal B}(\jp)$ & $10^4 {\cal B}(\Psi')$ &$\kappa$
                                                 \\ \hline\hline
$p \overline p$                 &21.4 $\pm$ 1.0&2.62 $\pm$ 0.57 &0.89 $\pm$ 0.20 \\
&         &1.9\phantom{0} $\pm$ 0.5\phantom{0} PDG\,& 0.64 $\pm$ 0.18 \\
$\Sigma^0\overline{\Sigma}{}^0$ &12.7 $\pm$ 1.7&1.20 $\pm$ 0.50  &0.63 $\pm$ 0.33 \\ 
$\Lambda  \overline \Lambda$    &13.5 $\pm$ 1.4& 1.89 $\pm$ 0.29    &0.99 $\pm$ 0.20 \\
$\Xi^- \overline {\Xi}{}^+$     &\phantom{1}9.0 $\pm$ 2.0&1.00 $\pm$ 0.32 &
                                             0.65 $\pm$ 0.35 \\  
$\Delta^{++}\overline{\Delta}{}^{--}$&11.0 $\pm$ 2.9&1.34 $\pm$ 0.35& 
                                             0.78 $\pm$ 0.28 \\ \hline
$\pi^+\pi^-$  &1.47\, $\pm$ 0.23&0.8 $\pm$ 0.5 PDG&4.25 $\pm$ 2.80 \\
$K^+ K^-$     &2.37 $\pm$ 0.31&1.0 $\pm$ 0.7 PDG&3.25 $\pm$ 2.30 \\ \hline
$\rho\pi$     &128\phantom{.} $\pm$ 10\phantom{.0}&$<$ 0.28 & $<$ 0.016  \\
              &           &$<$ 0.83 PDG$\,$ & $<$ 0.048  \\
$K^+\bar K^{*-} + c.c.$& 50.0 $\pm$ 4.0\phantom{0}&$<$ 0.3& $<$ 0.042 \\
                       &               &$<$ 0.54 PDG$\,$& $<$ 0.075 \\
$K^{0}\bar K^{*0} + c.c.$& 42.0 $\pm$ 4.0\phantom{0}& 0.81 $\pm$ 0.29&
                                                        0.13 $\pm$ 0.06 \\ 
$\omega\eta$  &15.8 $\pm$ 1.6\phantom{0}   &$<$ 0.33       & $<$ 0.15       \\
$\omega\eta'$ &1.67 $\pm$ 0.25  &0.76 $\pm$ 0.48& 2.94 $\pm$ 1.90\\
$\phi\eta$    &6.5\phantom{0} $\pm$ 0.7\phantom{0}&0.35 $\pm$ 0.20& 0.36 $\pm$ 0.21\\
$\phi\eta'$   &3.3\phantom{0} $\pm$ 0.4\phantom{0} &$<$ 0.75       & $<$ 1.35       \\[0.75em]
$\rho\eta$    &1.93 $\pm$ 0.23  &0.21 $\pm$ 0.12& 0.77 $\pm$ 0.45 \\
$\rho\eta'$   &1.05 $\pm$ 0.18  &$<$ 0.3        & $<$ 1.84       \\
$\omega\pi$   &4.2\phantom{0} $\pm$ 0.6\phantom{0}&0.38 $\pm$ 0.20& 0.67 $\pm$ 0.37\\
\hline
$K^{*0} K^{*0}$     & $<5.0$  & $0.45\pm 0.26$ & $>$ 0.66  \\ \hline

$\omega f_2$       & 43.0 $\pm$ 6.0  & $<$ 1.7  & $<$ 0.28        \\
$\rho a_2$         & 109\phantom{.} $\pm$ 22\phantom{.} & $<$ 2.3 & $<$ 0.15 \\
$K^{*0}\bar K^{*0}_2 + c.c.$& 67.0 $\pm$ 26\phantom{.} &$<$ 1.2 & $<$ 0.12   \\ 
$\phi f_2'$         & \phantom{0}8.0 $\pm$ 4.0 & $<$ 0.45 & $<$ 0.45 \\\hline
$b_1\pi + c.c.$    &30.0 $\pm$ 5.0& 5.3 $\pm$ 1.2 &1.28 $\pm$ 0.38\\
$K_1(1270)K^- + c.c.$& $<$ 29 BES\, & 10.0 $\pm$ 5.0 & $>$ 2.48\\
$K_1(1400)K^- + c.c.$& 38.0 $\pm$ 9.5 & $<$ 2.9& $<0.54$  \\ \hline       
\end{tabular}
\end{center}
\caption[]{Exclusive $\jp$ and $\psi'$ branching ratios and their ratios
 scaled by the corresponding electronic branching ratios and phase
 space corrected according to
 Eq.~(\ref{kappa}). Data are taken {}from \cite{PDG} in the $\jp$ cases
 and {}from the BES collaboration \cite{BES,bai99a,bai98c,Bai:1998fp} if not
 stated otherwise. Some of the BES data are preliminary; the BES data on
$\Psi'\to \rho\eta^({}'{}^), \phi\eta^({}'{}^)$ are quoted in Ref.\
\cite{che99}.} 
\label{tab:kappa}
\end{table}
%%%%%%%%%%%%%%%%%%%%%%%%%%%%%%%%%%%%%%%%%%%%%%%%%%%%%%%%%%%%%%%%%%%%%%%%%%%%%%%%%%

\begin{figure}[btp]
\begin{center}
\psfig{file=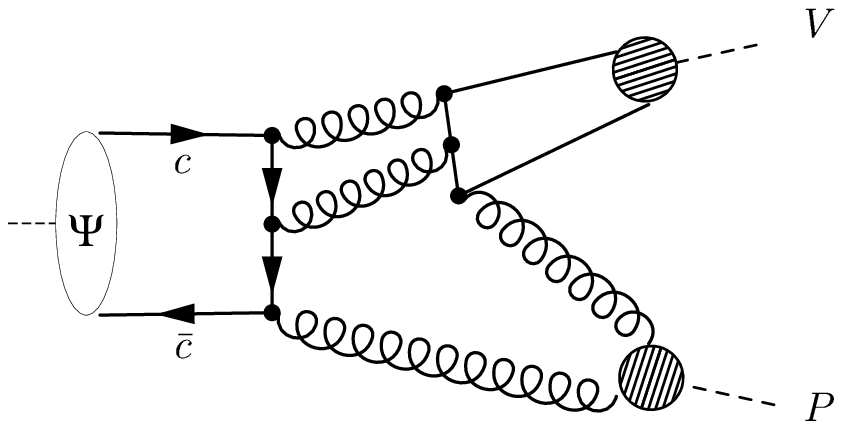, bb=60 640 310 775, height=4cm}
\end{center}
\caption{Schematic picture of 
the anomalous (doubly OZI-rule violating) decay mechanism for  
$\jp$ and $\Psi'$ decays into $PV$ meson pairs.}
\label{fig:ano} 
\end{figure}

The electromagnetic decay mechanism,
proceeding through the subprocess $\cbc\to \gamma^*\to h_1 h_2$ (see
Fig.~\ref{fig:elm}), also probes the charmonium wave function{} at small spatial
separation and, hence, would lead to $\kappa\simeq 1$ if it
dominates. This is what we observe, or at least what is
indicated, by the isospin-violating $\pi\omega$, $\eta\rho$ and
$\eta'\rho$ channels. For the other $\jp (\Psi')\to PV$ decays a
superposition of the mixing and the electromagnetic mechanism
occurs. For the flavor-singlet channels one has to consider  
an anomalous (formally doubly OZI-rule violating) contribution
\cite{sei88,bra97} too, see Fig.~\ref{fig:ano}. 
The mixing mechanism also applies to the  $\eta_{\c}\to \VbV$ decays 
with vector meson mixing replaced by that for
pseudoscalar mesons~\cite{FKS1}. 
%%%%%%%%%%%%%%%%%%%%%%%%%%%%%%%%%%%%%%%%%%%%%%%%%%%%%%%%%%%%%%%%%%%%%%%%%%
%%%%%%%%%%%%%%%%%%%%%%%%%%%%%%%%%%%%%%%%%%%%%%%%%%%%%%%%%%%%%%%%%%%%%%%%%%

The $\chi_{\c J}$ may also decay through light-quark Fock
components. However, in terms of meson mixing, the model would require
the coupling of appropriate $0^{++}$, $1^{++}$, $2^{++}$ mesons to
pairs of ground state mesons. These couplings are expected to be small
and, therefore, the $\chi_{\c J}\to \PbP, \VbV$ ($J=0,2$) decays seem
to be dominated by perturbative contributions while the hadronic
helicity non-conserving decays $\chi_{\c 1} \to \VbV$ should be
strongly suppressed. Indeed, the latter decays have not
been observed experimentally as yet \cite{PDG}. As we already mentioned in the
introduction the perturbative analysis \cite{BKS1,BKS2} of the
$\chi_{\c J}$ decays into the $\PbP$ channels provides results in fair
agreement with experiment \cite{PDG,bai98b}. In this analysis,  it is
however important to include the color-octet contributions, i.e.\ the
$\cbc g$ Fock state of the $\chi_{\c J}$. The contribution of the
valence Fock state alone, evaluated from the asymptotic form of the
pseudoscalar meson's light-cone wave function (as determined from the
analysis of the $P\gamma$ transition form factors \cite{JKR96,KR,FK}), is
substantially smaller than experiment. Aspects of the $\eta - \eta'$
mixing in these decays have been discussed in Ref.~\cite{FKS2}. The
extension of the perturbative analysis to the $\VbV$ channels, for
which first data exist \cite{bai98b}, is straightforward but the
required wave functions of the vector mesons are not known to a
sufficient degree of accuracy at present. 

%%%%%%%%%%%%%%%%%%%%%%%%%%%%%%%%%%%%%%%%%%%%%%%%%%%%%%%%%%%%%%%%%%%%%%%%%
\section{Mixing mechanism and the decays $\jp(\Psi') \to PV$} 
\label{sec:3}
%%%%%%%%%%%%%%%%%%%%%%%%%%%%%%%%%%%%%%%%%%%%%%%%%%%%%%%%%%%%%%%%%%%%

Let us now turn to the ${\jp}$ and $\Psi'$ decays into a pseudoscalar
and a vector meson. According to the discussion presented in 
Sec.~\ref{sec:2} we take into
account the three mechanisms shown in Figs.~\ref{fig:elm},
\ref{fig:strong} and \ref{fig:ano}. Each of these contributions will
be parametrized by a reduced amplitude, common to all $PV$ channels,
multiplied by flavor factors, meson-mixing factors and
flavor-symmetry corrections. Because of the lack of a compelling theory
we have to treat the reduced amplitudes as free parameters to be
adjusted to experiment. Despite this fact our ansatz which is
structurally similar, although different in detail, to those discussed
in Refs.~\cite{chen98,tua99,sei88,bra97} provides a systematic 
description of the 
${\jp}$ and $\Psi'$ decays into $PV$ mesons as we will see shortly. 

To begin with we discuss the electromagnetic decay mechanism shown in
Fig.~\ref{fig:elm}. Decomposing the helicity amplitudes for a decay
of an $n{}^3S_1$ charmonium state into $PV$ covariantly as 
\begin{eqnarray}
%&& 
{\cal M}_{0\lambda_V,\, \lambda}(n{}^3{\rm S}_1 \to VP) 
%\cr
%&& \quad = \
&=&  
  \frac{A_{PV}(n{}^3{\rm S}_1)}{M^2(n{}^3{\rm S}_1)}\, 
  \epsilon (p_1,p_2,\vep_V^*(\lambda_V),\vep(\lambda)) \ ,
\label{jp:pv}
\end{eqnarray}
we find for the electromagnetic contribution to the invariant
amplitude 
\begin{eqnarray}
%&& 
A_{PV}^{{\rm em}} (n{}^3{\rm S}_1) 
%\cr && \quad =  \ 
&=& 
4\pi\alpha_{\rm em} \, e_{\c} \, 
                e^{i\theta_e} \, f(n{}^3{\rm S}_1) \, 
                M(n{}^3{\rm S}_1) \; F_{PV}(s) \ .
\label{spv:elm}
\end{eqnarray}  
$f(n^3{\rm S}_1)$ is the decay constant of the $n^3{\rm S}_1$
charmonium state. Its numerical value is taken from Ref.~\cite{neu97}
 ($f(\jp) =405 ~\mev$ and $f(\Psi')=282 ~\mev$). $F_{PV}$ is
the time-like $P\to V$ transition form factor 
for transversely polarized vector mesons at $s=M^2(n{}^3{\rm S}_1)$. 
We write this form factor as  
\begin{equation}
F_{PV}(s) \;=\; c_{PV}^{\rm em}\, F_{\rho^0\pi^0}(s)\ ,
\label{eq:ff}
\end{equation}
and treat $F_{\rho^0\pi^0}(s)$ as a free parameter 
($c_{\rho^0\pi^0}^{\rm em}=1$ by definition). $c_{PV}^{\rm em}$
depends on the electric charges of the $PV$ system in units of the
elementary charge $e_0$. We also absorb $\eta-\eta'$ and $\omega-\phi$ 
mixing factors as well as a flavor symmetry breaking correction
factor, $y^{\rm em}$, into $c_{PV}^{{\rm em}}$.
The parameter $y^{\rm em}$
takes into account the difference between the occurrence of
strange and non-strange quarks in the form factors; its value will be
estimated below.
The  mixing of the pseudoscalar mesons is described in the quark-flavor 
scheme advocated in Ref.~\cite{FKS1}. Vector meson mixing is
treated analogously, the details are presented in the Appendix. 
Since the vector meson mixing angle, $\phi_V$, is so small 
(see Eq.~(\ref{mix-val})) we put $\cos{\phi_V}\simeq 1$ and 
$\sin{\phi_V}\simeq \phi_V$. The flavor factors $c_{PV}^{{\rm em}}$
are compiled in Table~\ref{tab:jpdec}. In Eq.~(\ref{spv:elm}) we also 
allow for a phase, $\theta_e$, relative to the strong interaction 
contributions discussed below. The transition form factors are
therefore to be understood as absolute values. The invariant function 
$A_{PV}$ in Eq.~(\ref{jp:pv}) is normalized in such a way that the decay 
width reads
\begin{equation}
\Gamma (n{}^3{\rm S}_1\to PV)\,=\, \frac{1}{96\pi}\,
\frac{\varrho_{PV}(n{}^3{\rm S}_1)^3}{M(n{}^3{\rm S}_1)} \, |A_{PV}|^2\,.
\end{equation}  
%%%%%%%%%%%%%%%%%%%%%%%%%%%%%%%%%%%%%%%%%%%%%%%%%%%%%%%%%%%%%%%%%
% TABLE 3
%%%%%%%%%%%%%%%%%%%%%%%%%%%%%%%%%%%%%%%%%%%%%%%%%%%%%%%%%%%%%%%%%%
\begin{table}[bt]
\begin{center}
\begin{tabular}{|c || c | c | c |}
\hline
$PV$ & 
$c_{PV}^{\rm em} $ & $c_{PV}^{\rm mix}$ & $c_{PV}^{\rm anom}$ \\
\hline\hline
$\omega\pi$    & $3$             & --  & --  \\
$\rho\eta$     & $3 \cos\phi_P$  & --  & --  \\
$\rho\eta'$    & $3 \sin\phi_P$  & --  &--   \\
$\phi\pi$      & $3 \phi_V$      & --  &--   \\
\hline 
$\rho^{0,\pm}\pi^{0,\mp}$  & $1 $ & $1$ & -- \\
$\omega\eta$     & $ \cos\phi_P$   & $\cos\phi_P$ &  $- \sqrt2 \, \tan\theta_8$ \\
$\omega\eta'$    & $\sin\phi_P$    & $\sin\phi_P$ &  $  \sqrt2$ \\
\hline
$K^{*+}K^{-}$    &$\phantom{-} y^{\rm em}$& $(\tilde y - 
                                     \sqrt2 \cot\tilde{\theta}_y)/2$ & -- \\
$K^{*0} \bar K^0$&$-2\, y^{\rm em}$& $(\tilde y - 
                                     \sqrt2 \cot\tilde{\theta}_y)/2 $& -- \\
$\phi \eta$      & $ \phantom{-} 2\sin\phi_P \, (y^{\rm em})^2$ 
                 & $\phantom{-}\sqrt{2}\sin\phi_P \, \cot\tilde{\theta}_y\,\tilde y$
                 &  $- \tilde y \tan\theta_8$ \\ 
$\phi \eta'$     & $-2\cos\phi_P\, (y^{\rm em})^2$ 
                 & $-\sqrt{2} \cos\phi_P\,\cot\tilde{\theta}_y\,\tilde y$
                 & $ \tilde y$ 
\\ \hline
\end{tabular}
\end{center}
\caption{The flavor factors appearing in the amplitudes for
$n{}^3S_1 \to PV$ decays. 
%($c_{\rho^0 \pi^0}^{\rm em} = c_{\rho^0 \pi^0}^{\rm mix}=1$). 
For definitions see 
Eqs.~(\protect\ref{eq:ff},\protect\ref{eq:mix},\protect\ref{eq:anom}). 
The parameters for $\eta-\eta'$ mixing,
$\phi_P$ and $\theta_8$, are defined in Ref.~\protect\cite{FKS1},
those describing vector meson mixing, $\phi_V$, $\tilde\theta_y$,
$\tilde{y}$, in the Appendix.}
\label{tab:jpdec}
\end{table}
%%%%%%%%%%%%%%%%%%%%%%%%%%%%%%%%%%%%%%%%%%%%%%%%%%%%%%%%%%%%%%%%%%%%%%%%%%%%%

The four isospin-violating channels now
allow a simple extraction of the $\rho^0\pi^0$ form factor. {}From a fit
to the experimental branching ratios of 
the isospin-violating decay channels  we find
\begin{eqnarray}
    F_{\rho^0\pi^0} (\mjp^2) &=& 27 \cdot 10^{-3}~{\rm GeV}^{-1} \ ,
    \cr
    F_{\rho^0\pi^0} (M_{\Psi'}^2) &=&  19 \cdot 10^{-3}~{\rm GeV}^{-1} \ .
\label{eq:ffres}
\end{eqnarray}
The results of the fit
are listed in Table~\ref{tab:ppdec}. 
 Note that the values (\ref{eq:ffres}) of the form factors are subject to
 errors which we estimate to 7.5\% and 21\% for the $\jp$ and $\Psi'$
 cases, respectively. 
The calculation of the time-like form factor, in particular in the
vicinity of resonances, is a delicate issue and, in so far, 
reliable theoretical information on this form factor is lacking.
Since the $\rho$ meson is transversely
polarized, the form factor is under control of higher-twist
contributions and should therefore fall off as $1/s^2$
asymptotically. Comparison of the two values in Eq.~(\ref{eq:ffres})
rather reveals an approximate $1/s$ behavior in the considered region of
$s$.
A recent QCD sum rule analysis \cite{kho97} of the
space-like $\rho\pi$ form factor yields values that are somewhat
smaller than those quoted in Eq.~(\ref{eq:ffres}). The enhancement of 
time-like form factors relative to space-like ones is known {}from the
pion and nucleon cases; the disparities typically amount to about a
factor of 3 in the charmonium region (see e.g.\
\cite{kro:93a,gou95}). In any case, measurements of the $PV$ transition
form factors in the charmonium region would be welcome. Such data
would permit a critical examination of the interpretation of the
electromagnetic contribution (\ref{spv:elm}) (cf.\ also the discussion
in Ref.~\cite{tua99}). In particular it might be checked whether or not
the neglect of contributions {}from the isospin-violating part of QCD
and {}from $\cbc$ annihilation mediated by $\gamma^* g^* g^*$ is indeed
justified.
  
The $\phi\pi$ channel offers direct possibility of testing
$\omega - \phi$ mixing. As inspection of Table~\ref{tab:ppdec} reveals
the predicted $\jp\to \phi\pi$ branching ratio obtained {}from our value
of the mixing angle (\ref{mix-val}), is well in agreement with the
experimental bound. A measurement of the branching ratio for the
$\phi\pi$ channel would allow a direct determination of $\phi_V$
through the ratio
\begin{equation}
\frac{{\cal B}(n^3S_1\to \phi\pi)}{{\cal B}(n^3S_1\to \omega\pi)}\,=\,
    \phi_V^2\, \left( \frac{\varrho_{\omega\pi}(n^3S_1)}
                         {\varrho_{\phi\pi}(n^3S_1)}\right)^3 \,,
\end{equation}
a relation that was proposed by Haber and Perrier \cite{hab85}
long time ago. 

%%%%%%%%%%%%%%%%%%%%%%%%%%%%%%%%%%%%%%%%%%%%%%%%%%%%%%%%%%%%%%%%%%%%%%%
% TABLE 4
%%%%%%%%%%%%%%%%%%%%%%%%%%%%%%%%%%%%%%%%%%%%%%%%%%%%%%%%%%%%%%%%%%%%%%%
\begin{table}[bt]
\begin{center}
\begin{tabular}{|c || c c | | c  c|}
\hline
$PV$ & $10^4\cdot{\cal B}_{\rm th}(\jp)$ &$10^4\cdot{\cal B}_{\rm exp}(\jp)$ & 
       $10^4\cdot{\cal B}_{\rm th}(\Psi')$ & $10^4\cdot{\cal B}_{\rm exp}(\Psi')$ \\
\hline\hline
$\omega\pi$ & $3.78$ % $4.58$ 
            & $4.20 \pm 0.60$ 
         &$0.38$ & $0.38 \pm 0.20$ \\
$\rho\eta$     & $2.04$ % $1.86$ 
               & $1.93 \pm 0.23$ 
         &$0.21$ & $ 0.21 \pm 0.12 $ \\
$\rho\eta'$     & $1.06$ % $0.96$ 
                & $1.05 \pm 0.18$  
         &$0.12$ & $ < 0.3  $  \\
$\phi\pi$     & $0.013$  & $< 0.068$ &$0.001$  & -- \\
\hline 
$\rho \pi $     & $127$ % $133$ 
                              & $128 \pm 10$ 
        &$0.13$ & $<0.28$ \\
$\omega\eta$     & $14.7$ % $15.3$ 
                 & $15.8 \pm 1.6$ 
                 &$ 0.12 $ & $ < 0.33$ \\
$\omega\eta'$     & $1.86$ % $1.69$ 
                  & $1.67 \pm 0.25$ 
                & $ 0.63 $ & $0.76 \pm 0.48$ \\
\hline
$K^{*+}K^{-}$ + c.c.  &$48 $  & $50 \pm 4$ 
                  &$ 0.13  $ & $< 0.30 $ \\
$K^{*0} \bar K^0$ + c.c. & $42$  & $42 \pm 4$ 
                   & $0.51 $ & $0.81 \pm 0.29$ \\
$\phi \eta$     & $7.5$  & $6.5 \pm 0.7$ 
                   &$0.16 $ & $0.35 \pm 0.20$ \\ 
$\phi \eta'$     & $2.6$  & $3.3  \pm 0.4$
                 &$0.46 $ & $< 0.75$
\\ \hline
\end{tabular}
\end{center}
\caption{Comparison of predictions and experiment for the branching
ratios (multiplied by $10^4$) for $\jp$ and $\Psi'$ decays into a
pseudoscalar and a vector meson.
The results for the first two sets of decay
   channels are obtained from fits (see text). The third set of decay
   channels involving strange mesons are predictions based on the
   parameters determined in the fits to the first two sets of data.
 The $\jp$ data are the PDG averages
\protect\cite{PDG}, the $\Psi'$ ones are preliminary BES data 
\protect\cite{BES,bai99a,che99}. The values of the mixing parameters for
vector mesons are quoted in the Appendix, those for pseudoscalar mesons are
$\phi_P=39.3^\circ$ and $\theta_8=-21.2^\circ$ \protect\cite{FKS1}.} 
\label{tab:ppdec}
\end{table}
%%%%%%%%%%%%%%%%%%%%%%%%%%%%%%%%%%%%%%%%%%%%%%%%%%%%%%%%%%%%%%%%%%%%

\vskip0.5em

The invariant amplitude for the strong decay mechanism (see
Fig.~\ref{fig:strong}) is parametrized as 
\begin{equation}
   A_{PV}^{\rm mix}(n{}^3S_1) \;=\; c_{PV}^{\rm mix} \, g^{\rm mix}(n{}^3S_1) \ ,
\label{eq:mix}
\end{equation}
where $ c_{PV}^{\rm mix}$ depends on flavor symmetry breaking and meson mixing
factors. As for the electromagnetic contribution we put
$c_{\rho_0\pi_0}^{\rm mix}=1$. By virtue of the node effect
(see the discussion in Sec.~\ref{sec:2})
this mechanism does not contribute to the $\Psi'$ decays,
i.e.\  we assume $g^{\rm mix}(\jp) \gg g^{\rm mix}({\Psi'})\simeq 0$ 
for the reduced amplitude. 

\vskip0.5em

The anomalous contribution, 
depicted in Fig.~\ref{fig:ano}, can only contribute to the flavor
singlet meson channels. It is assumed to proceed via a short-distance
annihilation of the initial $\cbc$ pair and a subsequent creation of
the $\eta$ and $\eta'$ states controlled by the matrix element
of the topological charge density
$\langle 0|\frac{\als}{4\pi} G \tilde{G}|\eta^{(}{}'{}^{)}\rangle$ 
which incorporates the $\rm U(1)_A$ anomaly. As shown in 
Refs.~\cite{FKS1,FKS2} the ratio of both the gluon matrix elements is
given by  
\begin{equation}
\frac{
\langle 0|\frac{\als}{4\pi} G \tilde{G}| \eta \rangle}{
\langle 0|\frac{\als}{4\pi} G \tilde{G}| \eta' \rangle}
\;=\; - \tan \theta_8 \ ,
\end{equation}
where the angle $\theta_8$ is related to the $\eta-\eta'$ mixing angle 
$\phi_P$ via $\theta_8 = \phi_P \, - \arctan[\sqrt2/y]$, 
where $y$ being the ratio of basic decay constants, $f_{\q}$ and
$f_{\s}$ in the quark-flavor scheme 
\cite{FKS1}. The numerical values for the mixing parameters,
determined in Ref.~\cite{FKS1}, read $\phi_P=39.3^\circ$,
$\theta_8=-21.2^\circ$ and $y=0.81$.
In view of these considerations we
parametrize the anomalous contribution as 
\begin{equation}
  A_{PV}^{\rm anom} (n{}^3S_1) \;=\; f(n{}^3S_1) \,  c_{PV}^{\rm
                             anom} \,  g^{\rm anom}({n{}^3S_1}) \ ,
\label{eq:anom}
\end{equation}
and define $c_{\omega\eta'}^{\rm anom}=\sqrt2$. The flavor factors 
$c_{PV}^{\rm anom}$ for the other channels can again be found
in Table~\ref{tab:jpdec}. Because of the short distance $\cbc$ annihilation
mechanism $g^{\rm anom}$, the reduced amplitude of the anomalous contribution,
should only mildly depend on the charmonium mass; we therefore assume
$g^{\rm anom}(\jp)=g^{\rm anom}(\Psi')$ for simplicity. 

\vskip0.5em

A fit to the ${\jp}$ and $\Psi'$ decay widths into $\rho\pi$,
$\omega\eta$ and $\omega\eta'$ provides the results shown in 
Table~\ref{tab:ppdec} for the following values of the parameters 
\begin{eqnarray}
&& g^{\rm mix}(\jp) = 0.020~{\rm GeV} \ , \cr
&& g^{\rm anom} = - 0.014 \ , \cr
&&\theta_e = 78^\circ \ .
\label{eq:values}
\end{eqnarray}
The value of the relative angle between the electromagnetic and the
strong contributions is very similar to that found by Bramon et 
al.~\cite{bra97}. Relative phases with values close
to $90^\circ$ between different decay amplitudes
have also been observed in other approaches to 
$J/\psi$ ($\psi'$) decays \cite{suz,bol98,Achasov:1999qj}.
Concerning the relative size of contributions from the
three individual decay mechanisms (electromagnetic, mixing,
anomalous), the result strongly depends on the considered decay
channel. For instance, in the case of $\jp \to \rho\pi$ the ratio of
$|A^{\rm em}|/|A^{\rm mix}|$ is only $0.1$ whereas for 
$\jp \to K^*{}^0 \bar K^0$ the same ratio is about $0.3$.
The anomalous contribution tends to interfere
with the mixing contribution destructively and is in some cases
very important. This is most drastically seen in
the channel $\jp \to \phi\eta'$ where the ratio 
$|A^{\rm em}|/|A^{\rm mix} + A^{\rm anom}|$ is about $1.1$.

\vskip0.5em

The decays into $KK^*$ and $\phi\eta(')$ involve strange quarks which
necessitates the consideration  of flavor symmetry breaking. In the case
of the electromagnetic contribution one has to pay attention to the
fact that the $PV$ transition form factors are controlled by higher-twist 
contributions since the vector mesons are transversely polarized. 
Hence, the form factors are proportional to a mass scale which, as 
shown by Efremov and Teryaev \cite{efr82}, is of the order of the 
hadron masses and not, as one may naively expect, set by current-quark 
masses. Modelling this mass scale by appropriate effective 
constituent-quark masses ($\hat{m}$), we estimate 
$y^{\rm em} \simeq \hat m_{\s}/\hat m_{\q} \simeq 1.3$. 
For $\phi\eta$ and $\phi\eta'$ channels the factor should appear
quadratically. For the mixing contribution strange quarks come either
{}from $\jp - \phi$ mixing, and are therefore $\propto \cot{\tilde
\theta_y}$, or by a soft creation of a $\sbs$ pair out of the
vacuum. This creation process is suppressed by a factor that is less
than unity, relative to the creation of a non-strange quark-antiquark 
pair. In accordance with the flavor symmetry breaking in vector meson
mixing, see the Appendix, we identify this factor with $\tilde y$
and use the value 0.8 for it. Note that in our mixing scheme 
$\sqrt2 \cot{\tilde\theta_y} = - \tilde y$, see Eq.~(\ref{mix-par}).
Therefore, the pattern of $\su3$ breaking is similar for the mixing
and the electromagnetic contributions.
With these values for the flavor symmetry factors and the
other already determined parameters at hand we are in the position to
predict the branching ratios for the decay channels involving
strange quarks. The results are also shown in Table~\ref{tab:ppdec}.
The quality of the predictions for the entire set of $\jp$ and
$\Psi'$ decays into a $PV$ pair is
good. For the $\jp$ decays the quality is similar to that obtained in
Ref.~\cite{bra97} although we have a smaller number of adjustable
parameters.

In our approach the $\Psi'$ decays into the $PV$ pair
electromagnetically with the exception of the flavor-singlet channels
where the anomalous mechanism contributes as well. In so far our
analysis differs {}from that advocated by Tuan \cite{tua99} who
extended the analysis in Ref.~\cite{bra97} to the case of the $\Psi'$. 
Tuan assumes the electromagnetic and the strong contributions
to cancel approximately in the $\Psi'\to\rho\pi$ channel. This 
assumption leads to substantially smaller branching ratios
for the $\Psi'\to\rho\pi$ and $\Psi'\to K^{*+} K^{-}$ channels
than we obtain. Since for both channels the BES collaboration only 
provides bounds the issue which of the contending models is correct, 
cannot be settled as yet. 

\vskip0.5em

Let us now dwell upon the compatibility of the fit value for $g(\jp)$
with the concept of $\jpomphi$ mixing. In this approach one
would write the reduced amplitude as 
\begin{equation}
g^{\rm mix}(\jp) \simeq \mjp^2 \,\tilde{\theta}_{\c}\, \sin{\tilde{\theta}_y}\, 
               g_{\omega\rho^0\pi^0} (s=\mjp^2)\,.
\label{eq:gesti}
\end{equation}
The $\omega\rho^0\pi^0$ vertex function at $s=0$ can be estimated from
chiral anomaly predictions for the coupling constants
$g_{P\gamma\gamma}$ and vector meson dominance\footnote{This 
approach has been used to predict the
radiative $PV$ transition, see \cite{fel99} and references therein.
 Kramer, Palmer and Pinsky 
investigated the $\omega\rho^0\pi^0$ vertex on the basis of effective
Lagrangians. Their numerical results are
in agreement with Eq.~(\ref{eq:vertex}). They also found
$|g_{\omega'\rho^0\pi^0}(s=0)| \ll |g_{\omega\rho^0\pi^0}(s=0)|$ which
is in line with our assumption $g^{{\rm mix}}(\Psi') \ll g^{{\rm
    mix}}(\jp)$.}:
\begin{equation}
     g_{\omega\rho^0\pi^0} (s=0) \,\simeq \,
                                             \frac{6 \, M_\rho \,
       M_\omega}{4 \pi^2 \, \sqrt2 \, f_\pi \, f_\rho \, f_\omega} \ .
\label{eq:vertex}
\end{equation}
Using our values for the $\jp - \omega$ mixing parameters
(\ref{mix-val}) and a monopole form for the $s$ dependence of the vertex
function, $\Lambda^2/(\mjp^2-\Lambda^2)$, we find that the fit value
of $g^{\rm mix}(\jp)$ (\ref{eq:values}) is reproduced for 
$\Lambda\simeq 1 ~\gev$. With regard to all the uncertainties
encountered in an estimate like this, a value of about 1 GeV for 
$\Lambda$ appears reasonable and is compatible with the typical
inverse radius of a light hadronic system.
Thus, we conclude that the small probability of the light-quark 
Fock components of the $\jp$ as estimated by $\jpomphi$ mixing, suffices to
generate the large $\jp\to PV$ branching ratios seen in experiment
since it is overcompensated by the very large soft $\omega, \phi \to PV$ 
transition.

%%%%%%%%%%%%%%%%%%%%%%%%%%%%%%%%%%%%%%%%%%%%%%%%%%%%%%%%%%%%%%%%%%%%%%%%%%
\section{Decays $\eta_{\c} \to \VbV$}
\label{sec:4}
%%%%%%%%%%%%%%%%%%%%%%%%%%%%%%%%%%%%%%%%%%%%%%%%%%%%%%%%%%%%%%%%%%%%%%%%%

We write the helicity amplitudes for the $\eta_{\c}\to \VbV$ decays
in a similar fashion as in Eq.~(\ref{jp:pv}) and parametrize the invariant
amplitude analogously to Eq.~(\ref{eq:mix}) as
\begin{equation}
  A_{\VbV}(\eta_{\c}) \; = \;  c_{\VbV}^{\rm mix} \, g^{\rm mix}(\eta_{\c}) \,,
\end{equation}
with the decay width given by
\begin{equation}
\Gamma (\eta_c \to \VbV)\,=\, \frac{1}{32\pi}\,
\frac{\varrho_{\VbV}(\eta_c)}{M_{\eta_c}} \, |A_{\VbV}|^2\,.
\end{equation}
We assume that the $\eta_{\c}$ decay is mediated by
the mixing of the $\eta_{\c}$ meson with the $\eta$ and $\eta'$.
The $\eta - \eta'$ mixing parameters 
have already been quoted. 
%In addition the ratio $y=f_q/f_s$ of basic decay constants in the
%quark-flavor mixing scheme~\cite{FKS1} enters.}
The mixing mechanism leads to the flavor factors $c_{\VbV}^{\rm
mix}$ shown in Table~\ref{tab:VV} ($c_{\rho^0\rho^0}^{\rm mix}
=1$). 
 Electromagnetic contributions to $\eta_{\c}$ decays
can only proceed via two photons and are thus
negligible (see also footnote~\ref{foot4}). 
Also a doubly OZI-rule violating decay mechanism
should be strongly suppressed here since the production of light
vector mesons through gluons is not enhanced by the $U(1)_A$ anomaly.
The possibility of a decay through the
$\cbc$ Fock component of the light mesons is, as in the case of the
$\jp$, ignored, see the discussion in Sec.~\ref{sec:2}.

%%%%%%%%%%%%%%%%%%%%%%%%%%%%%%%%%%%%%%%%%%%%%%%%%%%%%%%%%%%%%%%%%%%%%%%%
% TABLE 5
%%%%%%%%%%%%%%%%%%%%%%%%%%%%%%%%%%%%%%%%%%%%%%%%%%%%%%%%%%%%%%%%%%%%%%%
\begin{table}
\begin{center}
\begin{tabular}{|c || c ||c | c||}
\hline
$\VbV$ & $ c_{\VbV}^{\rm mix}$ &
         $10^3\cdot{\cal B}_{\rm th}$ & $10^3\cdot{\cal B}_{\rm exp}$ \\
\hline
\hline
$\rho\rho $ & $1 $ &    $16$ & $25 \pm 8 $ \cite{DM2} \\
$\omega\omega $ & $1$ &
         $5.2$  & $< 3.1$ \cite{MRK3a} \\
 & & & $<6.3$ \cite{DM2}\\
\hline
$K^* {\overline {K}}^*$  
  & $ (y + \sqrt2 \cot(\phi_P-\theta_8) )/2 $ 
    & $11.1$ & $8.2\pm 3.9$ \cite{MRK3}\\
  & & & $9.0 \pm 5.0$ \cite{MRK3a}\\
$\phi\phi $ 
  & $ y \, \sqrt2 \cot(\phi_P-\theta_8) $ 
    & $1.4$ & $7.1 \pm 2.2$ \cite{MRK3} \\
 & & & $3.1 \pm 1.2$ \cite{DM2}
\\ \hline
\end{tabular}
\end{center}
\caption{Flavor factors (quoted for single channels) of decay
amplitudes as well as experimental and theoretical branching ratios for
${\eta_{\c}} \to \VbV$ summed over all charge states. The quoted
theoretical values correspond to a fit with $g^{\rm mix}(\eta_{\c}) 
= 0.26~$GeV; the mixing parameters for the pseudoscalar mesons are 
$\phi_P=39.3^\circ$, $\theta_8=-21.2^\circ$ and $y=0.81$ \protect\cite{FKS1}.} 
\label{tab:VV}\end{table}
%%%%%%%%%%%%%%%%%%%%%%%%%%%%%%%%%%%%%%%%%%%%%%%%%%%%%%%%%%%%%%%%%%%%%%%%%

A fit to the experimental branching ratios~\footnote{Since the
two results for the $\phi\phi$ channel
quoted in Refs.~\cite{DM2,MRK3} are not compatible with each other,
we have doubled the respective errors in the fit.} 
\cite{DM2,MRK3a,MRK3}
yields $ g^{\rm mix}(\eta_{\c}) =
0.26~$GeV. The corresponding results are listed in Table~\ref{tab:VV}. 
The strength of $g^{\rm mix}(\eta_{\c})$ is compatible with the concept
of mixing. To see this we first note that here the same vertex function,
$g_{PVV},$ enters as in the $J/\psi\to PV$ decays in the
vector meson mixing approach. Analogously to 
Eqs.~(\ref{eq:gesti},\ref{eq:vertex}) the reduced amplitude reads
\begin{equation}
  g^{\rm mix}(\eta_{\c}) \, \simeq\, M^2_{\eta_{\c}}\,          
                        |\theta_{\c}|\, \sin{(\phi_P-\theta_8)}\,
                             g_{\eta_q\rho^0\rho^0}
                             (s=M^2_{\eta_{\c}}) \,,      
\label{eq:mix-eta}
\end{equation}
and the vertex function at $s=0$ is estimated by 
\begin{equation}
g_{\eta_q\rho^0\rho^0} (s=0) \,\simeq \,
  \frac{6 \, M_\rho^2}{4 \pi^2 \, \sqrt2 \, f_q \, f_\rho^2} \ .
\label{eq:eta0}
\end{equation}
Inserting the value $-1.0^\circ$ for the mixing angle
$\theta_c$ that controls the light quark admixture to the
$\eta_c$ as well as $f_q=1.07 \, f_\pi$~\cite{FKS1} 
and assuming again a monopole
behavior of the vertex function the value
$g^{\rm mix}(\eta_{\c})=0.26~\gev$ is reproduced for
$\Lambda \simeq 1.3$~GeV. This indicates a milder $s$ dependence of 
the vertex function in Eq.~(\ref{eq:mix-eta}) than in Eq.~(\ref{eq:gesti}).
We stress that the mixing approach explains the order of magnitude 
of the $\eta_{\c}\to \VbV$ decay widths correctly. This is, to our 
opinion, a highly non-trivial fact. In tendency our results agree 
better with the experimental results quoted by the DM-2 collaboration 
\cite{DM2} than with the ones found by MARK-III \cite{MRK3a,MRK3}. 
Better data for the $\eta_{\c}\to \VbV$ channels are obviously needed. 
In analogy to the $\psi'\to PV$ decays the $\eta_{\c}'\to \VbV$ decays are
expected to be strongly suppressed.

%%%%%%%%%%%%%%%%%%%%%%%%%%%%%%%%%%%%%%%%%%%%%%%%%%%%%%%%%%%%%%%%
\section{Comments on decays into baryon-antibaryon pairs}
%%%%%%%%%%%%%%%%%%%%%%%%%%%%%%%%%%%%%%%%%%%%%%%%%%%%%%%%%%%%%%%%
\label{sec:5}

A perturbative calculation of the $\jp$ and $\Psi'$ decays into
baryon-antibaryon pairs has been carried through in Ref.~\cite{bol98}
using the light-cone wave function for baryons proposed in
Ref.~\cite{BK96}. 
In contrast to
the $\chi_{\c J}$ decays here it suffices to consider the
color-singlet contributions only. The color-octet contributions are
suppressed by powers of $1/\mc$ and start  
at the order of the charm-quark velocity, $v$, squared. Moreover, there 
is no obvious enhancement 
of the corresponding hard scattering amplitudes which appear with at 
least the same power of $\als$ as the valence Fock state contributions. 
Thus, despite  the fact that $\mc$ is not very large and $v$ not
small ($v^2\simeq 0.3$), it seems reasonable to expect small higher 
Fock state contributions to both these decays. One may, however,
wonder what the strength of the mixing mechanism is. Since this mechanism is
responsible for the large $PV$ branching ratios 
(see Sec.~\ref{sec:2}) 
it may also contribute to the baryon-antibaryon channels
substantially and spoil the results presented in Ref.~\cite{bol98}. 
An estimate of the mixing contribution in analogy 
to Eqs.~(\ref{eq:gesti},\ref{eq:vertex}), using, for instance, the
$\omega$-nucleon coupling constants $g^{2}_{V\omega\pbp}/4\pi =
8.1\pm 1.5$ and
$g^{2}_{T\omega\pbp}/4\pi = 0.14\pm 0.2$ \cite{gre81}, reveals that it amounts
to less than $20\%$ of the experimental width if the $s$ dependence of
the $\omega\pbp$ vertex function is at least as strong as that of the
$\omega\rho\pi$ one. Since the $s$ dependence of a baryon vertex is
likely to be stronger than that of a meson vertex the mixing contribution
to the baryon-antibaryon channels is probably very small. Therefore 
the $\jp$ and $\Psi'$ decays into
baryon-antibaryon pairs seem to be under control of perturbative QCD. Indeed,
the order $v^0$ perturbative results obtained in Ref.~\cite{bol98} 
agree well with experiment
\cite{PDG,BES}. As we emphasized in Sec.~\ref{sec:2} the dominance
of the perturbative contribution is further supported by the fact that
the $\Psi'-\jp$ ratio $\kappa_{\BbB}$, defined in
Eq.~(\ref{kappa}), is in agreement with unity 
(see Table~\ref{tab:kappa}) 
as is indicative for short-distance $\cbc$ annihilation.

The hadronic helicity non-conserving decay $\eta_{\c}\to
\pbp$ may be estimated through $\eta_{\c} - \eta - \eta'$ mixing.
Using the coupling constant $g^2_{\eta N\overline{N}}/4\pi \simeq 1$ 
and $g_{\eta' N\overline{N}} \simeq 0$ \cite{fel99,gre81} and equating 
the $s$ dependence of the $\eta N\overline{N}$ vertex with that of the 
$\eta\VbV$ one as determined from the fit discussed in
Sec.~\ref{sec:4}, 
we find 
${\cal B} (\eta_{\c}\to\pbp) \simeq 0.5 \cdot 10^{-3}$. Comparing with
the experimental value of $(1.2\pm 0.4)\cdot 10^{-3}$ \cite{PDG} we
are tempted to conclude that the mixing mechanism also controls the  
$\eta_{\c}\to \pbp$ decays. 

%%%%%%%%%%%%%%%%%%%%%%%%%%%%%%%%%%%%%%%%%%%%%%%%%%%%%%%%%%%%%%%%
\section{Concluding remarks}
%%%%%%%%%%%%%%%%%%%%%%%%%%%%%%%%%%%%%%%%%%%%%%%%%%%%%%%%%%%%%%%%
\label{sec:sum}

Exclusive charmonium decays constitute a laboratory 
for investigating power corrections and higher Fock state
contributions as well as for studying the interplay of pQCD and 
soft mechanisms. That there are two distinct dynamical mechanisms at 
work can most easily be seen by comparison of $\jp\to PV$ and $\Psi'\to PV$
decays. Their scaling with the charmonium decay constants is indicative
for a short-distance mechanism (as, for instance, $\cbc$ annihilation
into virtual gluons or photons) while a clear deviation from that
scaling (as is seen for the helicity non-conserving $PV$ channels)
signals the prominent role of a more peripheral soft mechanism.
Several of the charmonium decays into pairs of pseudoscalar and/or
vector mesons are forbidden by spin and parity invariance. Others, for
instance the decays of the $C$-even charmonia into $PV$ channels, are
under control of the electromagnetic decay mechanism because  
either $G$-parity or isospin conservation is violated for the
channels involving non-strange mesons. Therefore, only the following
classes of decays are actually possible: the helicity non-conserving
$\jp (\Psi')\to PV$ and $\eta_{\c}\to \VbV$ decays which are not 
dominated by leading-twist pQCD; and the $\chi_{\c J}\to \PbP, \VbV$
($J=0,2$) decays which are accessible to a perturbative treatment and
the $\jp (\Psi')\to \PbP, \VbV$ decays. The latter decays are of
electromagnetic nature, the amplitudes being related to the time-like
$\PbP$ or $\VbV$ electromagnetic form factors at $s=M^2(n^3{\rm
S}_1)$. One should, however, be aware of corrections to the
electromagnetic decay mechanism from the isospin-violating part of
pQCD. In principle, allowed by strong interactions are also the
helicity non-conserving $\chi_{\c 1}\to \VbV$ decays. We however
expect very small branching ratios for these channels. The decays of 
the $C$-even charmonia into the $PV$ channels proceed through $\cbc$ 
annihilation into two photons and are therefore strongly suppressed. 
In fact these decays as well as the $\chi_{\c 1}\to \VbV$ have not 
been observed in experiment as yet \cite{PDG}.

The perturbative analysis of the $\chi_{\c J}\to \PbP$ decays has been
carried through in Refs.~\cite{BKS1,BKS2} and the results have been
found to agree fairly well with experiment although, with regard to
the recent BES data \cite{bai98b}, some readjustment of the wave
function parameters seems to be advisable. The $\VbV$ channels, on the
other hand, have not been analyzed as yet. 
Note that the required light-cone
wave functions of the vector mesons are not known to a sufficient degree
of accuracy. 

In the present paper we focus on the $\jp$ and $\Psi'$ 
decays into the $PV$ channels as well as the $\eta_{\c}\to\VbV$ decays.
We assume that in these cases the charmonium state
decays dominantly by a soft mechanism through Fock
components built up from light quarks only. This mechanism probes all
quark-antiquark separations in the charmonium wave function{}s and therefore
feels the difference between a $1{\rm S}$ and $2 {\rm S}$-wave
function. We model this mechanism by $\jpomphi$
and $\eta_{\c} - \eta - \eta'$ mixing, respectively, and argue
that this contribution is negligible for $\Psi'$ decays. 
The light vector and pseudoscalar
mesons couple through vertex functions to the final state mesons. With
a few parameters adjusted to experiment we find fair agreement with
the many channels measured for $\jp (\Psi')\to PV$ and
$\eta_{\c}\to\VbV$ decays. Most of the parameters
have a simple, direct physical interpretation and their magnitudes can
be estimated from physical considerations. 
Data of improved quality, in particular for the $\Psi'$ and
$\eta_{\c}$ decays are required for a more stringent test of our
approach. A richer database for the vector-tensor channels in
$\jp$ and $\Psi'$ decays would also allow the extension of our mixing
approach to this class of reactions.

\section*{Acknowledgments}

We thank Thomas Teubner for a critical reading of the manuscript.

\appendix

\section*{Appendix: Vector meson mixing}
%%%%%%%%%%%%%%%%%%%%%%%%%%%%%%%%%%%%%%%%%%%%%%%%%%%%%%%%%%%%%%%%%%%%%%%%%%
%\label{appA}
%

In order to estimate the strength of the light-quark Fock components in the
$\jp$ we consider $\jpomphi$ mixing in parallel to the
mixing of the pseudoscalar mesons. As in Refs.~\cite{FKS1,FKS2} (see
also \cite{fel99}) we start {}from the quark-flavor basis formed by the
three states $\omega_{\q}$, $\omega_{\s}$ and $\omega_{\c}$ which are
assumed to have the parton decompositions
\begin{eqnarray}
\label{basis}                                                     
 \mid \omega_{\q} \rangle &=& \Psi_{\q} \mid \ubu + \dbd \rangle
                                                      /\sqrt{2} +
                                                      \cdots \ , \nonumber\\
 \mid \omega_{\s} \rangle &=& \Psi_{\s} \mid \sbs \rangle + \cdots \ , \\
 \mid \omega_{\c} \rangle &=& \Psi_{\c} \mid \cbc \rangle + \cdots \nonumber
\end{eqnarray}
where the ellipses stand for higher Fock states. 
The physical meson states are related to the basis (\ref{basis}) by an
orthogonal transformation
\begin{eqnarray}
 \left( \begin{array}{c} \omega \\ \phi \\ \jp  \end{array} \right) \,=\, 
                  U(\phi_V,\tilde{\theta}_y,\tilde{\theta}_{\c})\;
              \left( \begin{array}{c} \omega_{\q} \\ \omega_{\s} 
                 \\ \omega_{\c} \end{array} \right)
\label{trans}
\end{eqnarray}
where 
\begin{eqnarray}
   U &=&
\left( \begin{array}{ccc}
  1  & - \phi_V  & - \tilde \theta_{\c} \sin\tilde \theta_y 
\\
  \phi_V & \phantom{-}1 & \phantom{-}\tilde \theta_{\c}\cos\tilde \theta_y 
\\
 \tilde \theta_{\c} \sin\tilde \theta_y 
     & -\tilde \theta_{\c} \cos\tilde \theta_y & 1 \end{array} \right) \,,
\label{uni} 
\end{eqnarray}
($U^\dagger U = 1 + {\cal O} (\tilde{\theta}_{\c}^2, \phi_V^2,
\tilde{\theta}_{\c}\phi_V)$). 
In Eq.~(\ref{uni}) $\tilde{\theta}_{\c}$ and $\phi_V$ are considered as
small quantities. $\tilde{\theta}_{\c}$ is small since the mixing
between the light and the heavy sector is an effect of ${\cal
O}(1/\mc^2)$ while the smallness of $\phi_V$ follows {}from the
well-known fact that the $\omega$ and the $\phi$ are nearly ideally
mixed \cite{sak67}.

The mass matrix in the quark-flavor basis is related to the
(diagonal) mass matrix for the physical mesons by
\begin{eqnarray}
%&& 
{\cal M}^2_{\q\s\c} 
% = \cr && 
&=&                 U^\dagger(\phi_V,\tilde{\theta}_y,\tilde{\theta}_{\c})\;
                     {\rm diag}[M^2_\omega, M^2_\phi, M^2_{\jp}]\;
                 U(\phi_V,\tilde{\theta}_y,\tilde{\theta}_{\c})\,.
%\cr &&
\label{mqsc}
\end{eqnarray}
Following our previous work on the pseudoscalar mesons
\cite{FKS1,FKS2}, we parametrize the mass matrix as 
\begin{eqnarray}
{\cal M}^2_{\q\s\c}
& = &
\left( \begin{array}{ccc}
  m_{qq}^2 + 2 \tilde a^2 
    & \tilde y \sqrt2 \tilde a^2 
      & \tilde z \sqrt2 \tilde a^2
\\
 \tilde y \sqrt2 \tilde a^2 
    & m_{ss}^2+ \tilde y^2 \tilde a^2 
        & \tilde y \tilde z \tilde a^2
\\
 \tilde z \sqrt2 \tilde a^2 
    & \tilde y \tilde z \tilde a^2 
        & m_{cc}^2 
\end{array}\right) \,,
\label{mij}
\end{eqnarray}
where $\tilde y$ and $\tilde z$ allow
for flavor symmetry breaking effects in the otherwise ``democratic''
mixing matrix whose strength, $\tilde a^2$, being controlled by 
quark-annihilation processes. 

Comparing the expressions (\ref{mqsc}) and (\ref{mij}) we find the
following six relations between the nine parameters
\begin{eqnarray}
&& m^2_{qq} + 2 \tilde a^2 = M^2_\omega\ , \cr 
&& m^2_{ss} + \tilde y^2 \tilde a^2 = M^2_\phi\ , \cr
&& m^2_{cc} = \mjp^2\ , \cr
&& \sqrt{2}\tilde y \tilde a^2 = (M^2_\phi - m^2_\omega) \phi_V \ , \cr
&& \tan{\tilde \theta_y} = -\sqrt{2}/\tilde y\ , \cr
&& \tilde \theta_{\c} = - \sqrt{2+\tilde y^2}\, \tilde z \tilde
a^2/\mjp^2 \ .
\label{mix-par}
\end{eqnarray}
One way to utilize these relations is to make use of flavor symmetry
in order to fix the quark mass terms. Thus, one may use
\begin{equation}
m^2_{qq} \simeq M^2_\rho \,, \quad \quad
m^2_{ss} \simeq 2 M^2_{K^*} - M^2_\rho \,,
\label{qmt}
\end{equation}
in analogy to the pseudoscalar meson case \cite{FKS1}. This would
allow the determination of the mixing parameters with the exception of
$\tilde z$. However, we dismiss this method since it leads to values
for the parameters which are instable under small corrections to the
relations (\ref{qmt}). 

An alternative method  that leads to robust values for the mixing
parameters, is to take the mixing angle $\phi_V$ {}from experiment and to
fix the parameters controlling flavor symmetry breaking in analogy to
the pseudoscalar case by
\begin{equation}
  \tilde y = f_\omega/f_\phi\,, \quad \quad \tilde z = f_\omega/f_{\jp}\,.
\label{fsb}
\end{equation}
This is an assumption here in contrast to the pseudoscalar case where
the divergences of the axial vector currents, embodying 
the $U(1)_A$ anomaly, relate the flavor symmetry breaking parameters
to the meson decay constants.
{}From the values of the decay constants quoted in
Ref.~\cite{neu97}, we obtain $\tilde y = 0.80 \pm 0.04$ and $\tilde z = 0.48
\pm 0.02$. The mixing ansatz (\ref{trans}) directly relates the ratio
of the widths for the radiative $\phi- \pi^0$ and $\omega - \pi^0$ transitions to
the mixing angle $\phi_V$ \cite{bra97}
\begin{equation}
\frac{\Gamma(\phi\to\pi^0\gamma)}{\Gamma(\omega\to\pi^0\gamma)} =
       \phi_V^2 \, \left[ \frac{M_\phi}{M_\omega} \, 
              \frac{\varrho_{\pi\gamma}(\phi)}{\varrho_{\pi\gamma}(\omega)}
                                   \right]^3 \,.
\end{equation}
{}From the experimental decay widths \cite{PDG} we find
\begin{equation}
\phi_V = 3.4^\circ \pm 0.2^\circ\,,
\label{mix-angle}
\end{equation}
where the sign is chosen according to Eq.~(\ref{mix-par}). This
result is consistent with other determinations of that angle \cite{sak67,ruj}. 

Using Eqs.~(\ref{fsb},\ref{mix-angle}) and of course the masses of the
physical vector mesons as input to Eq.~(\ref{mix-par}), we obtain for the 
other mixing parameters
\begin{eqnarray}
&& \tilde a^2 = 0.022 \pm 0.002 ~\gev^2\ , \cr
&& \tilde \theta_y = - 59.8^\circ\,\pm 1.3^\circ\ , \cr
&& \tilde \theta_{\c} = - 0.10^\circ \pm 0.01^\circ \ .
\label{mix-val}
\end{eqnarray}
The mixing strength,
$\tilde a^2$ is much smaller than in the case of the pseudoscalar
mesons \cite{FKS1} where it is dominated by the $U(1)_A$ anomaly.
For completeness we note that an evaluation of the quark mass terms
{}from these values for the mixing parameters leads to values which
merely deviate by about $2\%$ {}from those obtained via the flavor
symmetry relations (\ref{qmt}).

Our approach leads to the following quark content of the physical
mesons
\begin{eqnarray}
| \omega \rangle &=& | \omega_{\q} \rangle 
            - 0.060 \, | \omega_{\s} \rangle 
            - 1.5 \cdot 10^{-3} \, | \omega_{\c} \rangle\ , \nonumber\\
| \phi \rangle &=&  | \omega _{\s} \rangle     
               + 0.060 \, | \omega_{\q} \rangle 
            - 0.9 \cdot 10^{-3} \,  | \omega_{\c} \rangle \ ,\nonumber\\
| \jp \rangle &=&  | \omega_{\c} \rangle  
                 + 1.5\cdot 10^{-3} \,  | \omega_{\q} \rangle 
                 + 0.9 \cdot 10^{-3} \, | \omega_{\s} \rangle 
                 \ .
\label{mixprops}
% \cr &&
\end{eqnarray}

%
%%%%%%%%%%%%%%%%%%%%%%%%%%%%%%%%%%%%%%%%%%%%%%%%%%%%%%%%%%%%%%%%%%%%%
%\begin{references}


\begin{thebibliography}{9}
%%%%%%%%%%%%%%%%%%%%%%%%%%%%%%%%%%%%%%%%%%%%%%%%%%%%%%%%%%%%%%%%%%%%


%\cite{Duncan:1980qd}
\bibitem{dun80}
A.~Duncan and A.~Mueller,
%``Heavy Quarkonium Decays And The Renormalization Group,''
Phys.\ Lett.\  {\bf B93}, 119 (1980).
%%CITATION = PHLTA,B93,119;%%
%\href{http://www.slac.stanford.edu/spires/find/hep/www?j=PHLTA%2cB93%2c119}{SPIRES}

%\cite{Brodsky:1981kj}
\bibitem{BrL81}
S.~J.~Brodsky and G.~P.~Lepage,
%``Helicity Selection Rules And Tests Of Gluon Spin In Exclusive QCD Processes,''
Phys.\ Rev.\  {\bf D24}, 2848 (1981).
%%CITATION = PHRVA,D24,2848;%%
%\href{http://www.slac.stanford.edu/spires/find/hep/www?j=PHRVA%2cD24%2c2848}{SPIRES}

%\cite{Chernyak:1982zz}
\bibitem{che82}
V.~L.~Chernyak and A.~R.~Zhitnitsky,
%``Exclusive Decays Of Heavy Mesons,''
Nucl.\ Phys.\  {\bf B201}, 492 (1982).
%%CITATION = NUPHA,B201,492;%%
%\href{http://www.slac.stanford.edu/spires/find/hep/www?j=NUPHA%2cB201%2c492}{SPIRES}

%\cite{Appelquist:1975zd}
\bibitem{app75}
T.~Appelquist and H.~D.~Politzer,
%``Orthocharmonium And E+ E- Annihilation,''
Phys.\ Rev.\ Lett.\  {\bf 34}, 43 (1975).
%%CITATION = PRLTA,34,43;%%
%\href{http://www.slac.stanford.edu/spires/find/hep/www?j=PRLTA%2c34%2c43}{SPIRES}

 
%\cite{Bolz:1998ez}
\bibitem{BKS1}
J.~Bolz, P.~Kroll and G.~A.~Schuler,
%``Higher Fock states and power counting in exclusive P-wave quarkonium  decays,''
Eur.\ Phys.\ J.\  {\bf C2}, 705 (1998)
[hep-ph/9704378].
%%CITATION = HEP-PH 9704378;%%
%\href{http://www.slac.stanford.edu/spires/find/hep/www?eprint=HEP-PH/9704378}{SPIRES}


%\cite{Bodwin:1995jh}
\bibitem{BBL}
G.~T.~Bodwin, E.~Braaten and G.~P.~Lepage,
%``Rigorous QCD analysis of inclusive annihilation and production of heavy quarkonium,''
Phys.\ Rev.\  {\bf D51}, 1125 (1995)
[hep-ph/9407339].
%%CITATION = HEP-PH 9407339;%%
%\href{http://www.slac.stanford.edu/spires/find/hep/www?eprint=HEP-PH/9407339}{SPIRES}

%\newcommand{\wwwspires}{http://www.slac.stanford.edu/spires/find/hep/www}
%\cite{Wong:2000hc}
\bibitem{wong}
S.~M.~Wong,
%``Colour octet contribution in exclusive P-wave charmonium decay into  octet and decuplet baryons,''
Eur.\ Phys.\ J.\  {\bf C14}, 643 (2000)
[hep-ph/9903236];
%%CITATION = HEP-PH 9903236;%%
%\href{\wwwspires?eprint=HEP-PH/9903236}{SPIRES}
%\cite{Wong:2000dj}
%``Colour octet contribution in exclusive P-wave charmonium decay into  proton antiproton,''
Nucl.\ Phys.\  {\bf A674}, 185 (2000)
[hep-ph/9903221].
%%CITATION = HEP-PH 9903221;%%
%\href{\wwwspires?eprint=HEP-PH/9903221}{SPIRES}

%\cite{Caso:1998tx}
\bibitem{PDG}
Particle Data Group: Review of Particle Properties
(C.~Caso {\it et al.}),
%``Review of particle physics,''
Eur.\ Phys.\ J.\  {\bf C3}, 1 (1998).
%%CITATION = EPHJA,C3,1;%%
%\href{http://www.slac.stanford.edu/spires/find/hep/www?j=EPHJA%2cC3%2c1}{SPIRES}

%\cite{Brodsky:1997fj}
\bibitem{bro97}
S.~J.~Brodsky and M.~Karliner,
%``Intrinsic charm of vector mesons: A possible solution of the *rho pi  puzzle*,''
Phys.\ Rev.\ Lett.\  {\bf 78}, 4682 (1997)
[hep-ph/9704379].
%%CITATION = HEP-PH 9704379;%%
%\href{http://www.slac.stanford.edu/spires/find/hep/www?eprint=HEP-PH/9704379}{SPIRES}

%\cite{Chen:1998ma}
\bibitem{chen98}
Y.~Chen and E.~Braaten,
%``An explanation for the rho pi puzzle of J/psi and psi' decays,''
Phys.\ Rev.\ Lett.\  {\bf 80}, 5060 (1998)
[hep-ph/9801226].
%%CITATION = HEP-PH 9801226;%%
%\href{http://www.slac.stanford.edu/spires/find/hep/www?eprint=HEP-PH/9801226}{SPIRES}

%\cite{Clavelli:1983rk}
\bibitem{cla83}
L.~J.~Clavelli and G.~W.~Intemann,
%``Vector Meson Mixing And Hadronic Decays Of Psi,''
Phys.\ Rev.\  {\bf D28}, 2767 (1983).
%%CITATION = PHRVA,D28,2767;%%
%\href{http://www.slac.stanford.edu/spires/find/hep/www?j=PHRVA%2cD28%2c2767}{SPIRES}

%\cite{Pinsky:1985da}
\bibitem{pin90}
S.~S.~Pinsky,
%``An Analysis Of The Decay Psi $\to$ Vector + Pseudoscalar,''
Phys.\ Rev.\  {\bf D31}, 1753 (1985)
%%CITATION = PHRVA,D31,1753;%%
%\href{http://www.slac.stanford.edu/spires/find/hep/www?j=PHRVA%2cD31%2c1753}{SPIRES}
and
%\cite{Pinsky:1990ue}
% S.~S.~Pinsky,
%``The Psi-Prime To J / Psi Hadronic Decay Puzzle,''
Phys.\ Lett.\  {\bf B236}, 479 (1990).
%%CITATION = PHLTA,B236,479;%%
%\href{http://www.slac.stanford.edu/spires/find/hep/www?j=PHLTA%2cB236%2c479}{SPIRES}

%\cite{Li:1997yn}
\bibitem{li97}
X.~Li, D.~V.~Bugg and B.~Zou,
%``A possible explanation of the *rho pi puzzle* in J/psi, psi' decays,''
Phys.\ Rev.\  {\bf D55}, 1421 (1997).
%%CITATION = PHRVA,D55,1421;%%
%\href{http://www.slac.stanford.edu/spires/find/hep/www?j=PHRVA%2cD55%2c1421}{SPIRES}


%\cite{Suzuki:1998ea}
\bibitem{suz}
M.~Suzuki,
%``Long-distance final-state interactions and J/psi decay,''
Phys.\ Rev.\  {\bf D57}, 5717 (1998)
[hep-ph/9801284];
%%CITATION = HEP-PH 9801284;%%
%\href{http://www.slac.stanford.edu/spires/find/hep/www?eprint=HEP-PH/9801284}{SPIRES}
%\cite{Suzuki:1999nb}
%M.~Suzuki,
%``A large final-state interaction in the 0- 0- decays of J/psi,''
Phys.\ Rev.\  {\bf D60}, 051501 (1999)
[hep-ph/9901327].
%%CITATION = HEP-PH 9901327;%%
%\href{\wwwspires?eprint=HEP-PH/9901327}{SPIRES}

%\cite{Hou:1983kh}
\bibitem{hou83}
W.~Hou and A.~Soni,
%``Vector Gluonium As A Possible Explanation For Anomalous Psi Decays,''
Phys.\ Rev.\ Lett.\  {\bf 50}, 569 (1983);
%%CITATION = PRLTA,50,569;%%
%\href{http://www.slac.stanford.edu/spires/find/hep/www?j=PRLTA%2c50%2c569}{SPIRES}
%
%\cite{Chan:1999px}
C.~Chan and W.~Hou,
%``On the mixing amplitude of J/psi and vector glueball O,''
Nucl.\ Phys.\  {\bf A675}, 367 (2000)
[hep-ph/9911423].
%%CITATION = HEP-PH 9911423;%%
%\href{http://www.slac.stanford.edu/spires/find/hep/www?eprint=HEP-PH/9911423}{SPIRES}

%\cite{Tuan:1999ig}
\bibitem{tua99}
S.~F.~Tuan,
%``The rho pi puzzle of J/psi and psi' decays,''
Commun.\ Theor.\ Phys.\  {\bf 33}, 285 (2000)
[hep-ph/9903332];
%%CITATION = HEP-PH 9903332;%%
%\href{http://www.slac.stanford.edu/spires/find/hep/www?eprint=HEP-PH/9903332}{SPIRES}
%\cite{Gu:1999uq}
Y.~F.~Gu and S.~F.~Tuan,
%``Charmonium decay physics,''
Nucl.\ Phys.\  {\bf A675}, 404 (2000)
[hep-ph/9910423].
%%CITATION = HEP-PH 9910423;%%
%\href{\wwwspires?eprint=HEP-PH/9910423}{SPIRES}
%\cite{Kroll:1993zx}
\bibitem{kro:93a}
P.~Kroll, T.~Pilsner, M.~Sch{\"u}rmann and W.~Schweiger,
%``On exclusive reactions in the timelike region,''
Phys.\ Lett.\  {\bf B316}, 546 (1993)
[hep-ph/9305251].
%%CITATION = HEP-PH 9305251;%%
%\href{http://www.slac.stanford.edu/spires/find/hep/www?eprint=HEP-PH/9305251}{SPIRES}

%\cite{Seiden:1988rr}
\bibitem{sei88}
A.~Seiden, H.~F.~Sadrozinski and H.~E.~Haber,
%``Doubly Ozi Violating Effects In J / Psi Decays,''
Phys.\ Rev.\  {\bf D38}, 824 (1988).
%%CITATION = PHRVA,D38,824;%%
%\href{http://www.slac.stanford.edu/spires/find/hep/www?j=PHRVA%2cD38%2c824}{SPIRES}

%\cite{Bramon:1997mf}
\bibitem{bra97}
A.~Bramon, R.~Escribano and M.~D.~Scadron,
%``Mixing of eta - eta' mesons in J/psi decays into a vector and a  pseudoscalar meson,''
Phys.\ Lett.\  {\bf B403}, 339 (1997)
[hep-ph/9703313].
%%CITATION = HEP-PH 9703313;%%
%\href{http://www.slac.stanford.edu/spires/find/hep/www?eprint=HEP-PH/9703313}{SPIRES}

%\cite{Feldmann:1998vh}
\bibitem{FKS1}
T.~Feldmann, P.~Kroll and B.~Stech,
%``Mixing and decay constants of pseudoscalar mesons,''
Phys.\ Rev.\  {\bf D58}, 114006 (1998)
[hep-ph/9802409].
%%CITATION = HEP-PH 9802409;%%
%\href{http://www.slac.stanford.edu/spires/find/hep/www?eprint=HEP-PH/9802409}{SPIRES}

%\cite{Harris:1999wn}
\bibitem{BES}
F.~A.~Harris,
%``Recent charmonium results from BES,''
hep-ex/9903036 and references therein.
%%CITATION = HEP-EX 9903036;%%
%\href{http://www.slac.stanford.edu/spires/find/hep/www?eprint=HEP-EX/9903036}{SPIRES}

\bibitem{CZ84}
V.~L.~Chernyak and A.~R.~Zhitnitsky,
%``Asymptotic Behavior Of Exclusive Processes In QCD,''
Phys.\ Rept.\  {\bf 112}, 173 (1984).
%%CITATION = PRPLC,112,173;%%
%\href{http://www.slac.stanford.edu/spires/find/hep/www?j=PRPLC%2c112%2c173}{SPIRES}

%\cite{Bolz:1997wh}
\bibitem{BKS2}
J.~Bolz, P.~Kroll and G.~A.~Schuler,
%``Colour-octet contributions to exclusive charmonium decays,''
Phys.\ Lett.\  {\bf B392}, 198 (1997)
[hep-ph/9610265].
%%CITATION = HEP-PH 9610265;%%
%\href{http://www.slac.stanford.edu/spires/find/hep/www?eprint=HEP-PH/9610265}{SPIRES}

%\cite{Bai:1999mq}
\bibitem{bai99a}
J.~Z.~Bai {\it et al.}  [BES Collaboration],
%``Charmonium decays to axialvector plus pseudoscalar mesons,''
Phys.\ Rev.\ Lett.\  {\bf 83}, 1918 (1999)
[hep-ex/9901022].
%%CITATION = HEP-EX 9901022;%%
%\href{http://www.slac.stanford.edu/spires/find/hep/www?eprint=HEP-EX/9901022}{SPIRES}

%\cite{Bai:1998fu}
\bibitem{bai98c}
J.~Z.~Bai {\it et al.}  [BES Collaboration],
%``Decays of the J/psi to Lambda Antilambda, Lambda Antilambda gamma and  Lambda Antilambda pi0 final states,''
Phys.\ Lett.\  {\bf B424}, 213 (1998);
%%CITATION = PHLTA,B424,213;%%
%\href{http://www.slac.stanford.edu/spires/find/hep/www?j=PHLTA%2cB424%2c213}{SPIRES}
Erratum {\it ibid.}\/ {\bf B438}, 447 (1998).

%\newcommand{\wwwspires}{http://www.slac.stanford.edu/spires/find/hep/www}
%\cite{Bai:1998fp}
\bibitem{Bai:1998fp}
J.~Z.~Bai {\it et al.}  [BES Collaboration],
%``psi(2S) hadronic decays to vector-tensor final states,''
Phys.\ Rev.\ Lett.\  {\bf 81}, 5080 (1998).
%%CITATION = PRLTA,81,5080;%%
%\href{\wwwspires?j=PRLTA%2c81%2c5080}{SPIRES}
%\cite{Chernyak:1999cj}

\bibitem{che99}
V.~Chernyak,
%``J/psi theory review, or from J to psi,''
hep-ph/9906387
%%CITATION = HEP-PH 9906387;%%
%\href{http://www.slac.stanford.edu/spires/find/hep/www?eprint=HEP-PH/9906387}{SPIRES}
[talk given at the
International Workshop on $e^+ e^-$ collisions from $\phi$ to $\jp$,
Novosibirsk (1999)] and references therein. 



\bibitem{nov78}
V.~A.~Novikov, L.~B.~Okun, M.~A.~Shifman, A.~I.~Vainshtein, M.~B.~Voloshin and V.~I.~Zakharov,
%``Charmonium Annihilation In Quantum Chromodynamics (Nonrelativistic Approach),''
Phys.\ Rept.\  {\bf 41}, 1 (1978).
%%CITATION = PRPLC,41,1;%%
%\href{http://www.slac.stanford.edu/spires/find/hep/www?j=PRPLC%2c41%2c1}{SPIRES}


\bibitem{mey79} H.~Meyer,
%``Measurements Of The Properties Of The Upsilon Family From E+ E- Annihilation,''
 Invited talk given at the 1979 Int. Symp. on Lepton and Photon Interactions
  at High Energies, Batavia, Ill., Aug 23-29, 1979 [DESY 79/81].

%\cite{Kopeliovich:1993gk}
\bibitem{pei98}
B.~Z.~Kopeliovich, J.~Nemchick, N.~N.~Nikolaev and B.~G.~Zakharov,
%``Novel color transparency effect: Scanning the wave function of vector mesons,''
Phys.\ Lett.\  {\bf B309}, 179 (1993)
[hep-ph/9305225];
%%CITATION = HEP-PH 9305225;%%
%\href{http://www.slac.stanford.edu/spires/find/hep/www?eprint=HEP-PH/9305225}{SPIRES}
%
%\cite{Hoyer:1999xe}
P.~Hoyer and S.\ Peign\'{e},
%``psi' to psi ratio in diffractive photoproduction,''
Phys.\ Rev.\ {\bf D61}, 031501 [hep-ph/9909519].
%%CITATION = HEP-PH 9909519;%%
%\href{http://www.slac.stanford.edu/spires/find/hep/www?eprint=HEP-PH/9909519}{SPIRES}

%\cite{Kramer:1984cy}
\bibitem{kra84}
G.~Kramer, W.~F.~Palmer and S.~S.~Pinsky,
%``Testing Chiral Anomalies With Hadronic Currents,''
Phys.\ Rev.\  {\bf D30}, 89 (1984).
%%CITATION = PHRVA,D30,89;%%
%\href{http://www.slac.stanford.edu/spires/find/hep/www?j=PHRVA%2cD30%2c89}{SPIRES}

%\cite{Bai:1996rd}
\bibitem{bai96}
J.~Z.~Bai {\it et al.}  [BES Collaboration],
%``Search for a vector glueball by a scan of the J / psi resonance,''
Phys.\ Rev.\  {\bf D54}, 1221 (1996).
%%CITATION = PHRVA,D54,1221;%%
%\href{http://www.slac.stanford.edu/spires/find/hep/www?j=PHRVA%2cD54%2c1221}{SPIRES}

%\cite{Bai:1999cw}
\bibitem{bai98b}
J.~Z.~Bai {\it et al.}  [BES Collaboration],
%``Study of the hadronic decays of chi/c states,''
Phys.\ Rev.\  {\bf D60}, 072001 (1999)
[hep-ex/9812016].
%%CITATION = HEP-EX 9812016;%%
%\href{http://www.slac.stanford.edu/spires/find/hep/www?eprint=HEP-EX/9812016}{SPIRES}

%\cite{Jakob:1996hd}
\bibitem{JKR96}
R.~Jakob, P.~Kroll and M.~Raulfs,
%``Meson - photon transition form-factors,''
J.\ Phys.\ {\bf G22}, 45 (1996)
[hep-ph/9410304].
%%CITATION = HEP-PH 9410304;%%
%\href{http://www.slac.stanford.edu/spires/find/hep/www?eprint=HEP-PH/9410304}{SPIRES}


%\cite{Kroll:1996jx}
\bibitem{KR}
P.~Kroll and M.~Raulfs,
%``The pi gamma transition form factor and the pion wave function,''
Phys.\ Lett.\  {\bf B387}, 848 (1996)
[hep-ph/9605264].
%%CITATION = HEP-PH 9605264;%%
%\href{http://www.slac.stanford.edu/spires/find/hep/www?eprint=HEP-PH/9605264}{SPIRES}

%\cite{Feldmann:1998vc}
\bibitem{FK}
T.~Feldmann and P.~Kroll,
%``Flavor symmetry breaking and mixing effects in the eta gamma and  eta' gamma transition form factors,''
Eur.\ Phys.\ J.\  {\bf C5}, 327 (1998)
[hep-ph/9711231].
%%CITATION = HEP-PH 9711231;%%
%\href{http://www.slac.stanford.edu/spires/find/hep/www?eprint=HEP-PH/9711231}{SPIRES}

%\cite{Feldmann:1999sh}
\bibitem{FKS2}
T.~Feldmann, P.~Kroll and B.~Stech,
%``Mixing and decay constants of pseudoscalar mesons: The sequel,''
Phys.\ Lett.\  {\bf B449}, 339 (1999)
[hep-ph/9812269].
%%CITATION = HEP-PH 9812269;%%
%\href{http://www.slac.stanford.edu/spires/find/hep/www?eprint=HEP-PH/9812269}{SPIRES}

%\cite{Neubert:1997uc}
\bibitem{neu97}
M.~Neubert and B.~Stech,
%``Non-leptonic weak decays of B mesons,''
 in Heavy Flavours II, Eds. A.J.\ Buras and M.\ Lindner,
                World Scientific, Singapore
[hep-ph/9705292].
%%CITATION = HEP-PH 9705292;%%
%\href{http://www.slac.stanford.edu/spires/find/hep/www?eprint=HEP-PH/9705292}{SPIRES}

%\cite{Khodjamirian:1999tk}
\bibitem{kho97}
A.~Khodjamirian,
%``Form factors of gamma* rho --> pi and gamma* gamma --> pi0 transitions  and light-cone sum rules,''
Eur.\ Phys.\ J.\  {\bf C6}, 477 (1999)
[hep-ph/9712451].
%%CITATION = HEP-PH 9712451;%%
%\href{http://www.slac.stanford.edu/spires/find/hep/www?eprint=HEP-PH/9712451}{SPIRES}

%\cite{Gousset:1995yh}
\bibitem{gou95}
T.~Gousset and B.~Pire,
%``Timelike form-factors at high-energy,''
Phys.\ Rev.\  {\bf D51}, 15 (1995)
[hep-ph/9403293];
%%CITATION = HEP-PH 9403293;%%
%\href{http://www.slac.stanford.edu/spires/find/hep/www?eprint=HEP-PH/9403293}{SPIRES}
%
%\cite{Jakob:1993iw}
R.~Jakob and P.~Kroll,
%``The Pion form-factor: Sudakov suppressions and intrinsic transverse momentum,''
Phys.\ Lett.\  {\bf B315}, 463 (1993)
[hep-ph/9306259]
%%CITATION = HEP-PH 9306259;%%
%\href{http://www.slac.stanford.edu/spires/find/hep/www?eprint=HEP-PH/9306259}{SPIRES}
Erratum {\it ibid.}\/ {\bf B319}, 545 (1993).


%\cite{Haber:1985cv}
\bibitem{hab85}
H.~E.~Haber and J.~Perrier,
%``A Model Independent Analysis Of Hadronic Decays Of J / Psi And Eta(C) (2980),''
Phys.\ Rev.\  {\bf D32}, 2961 (1985).
%%CITATION = PHRVA,D32,2961;%%
%\href{http://www.slac.stanford.edu/spires/find/hep/www?j=PHRVA%2cD32%2c2961}{SPIRES}

%\cite{Bolz:1998as}
\bibitem{bol98}
J.~Bolz and P.~Kroll,
%``Exclusive J/psi and psi' decays into baryon antibaryon pairs,''
Eur.\ Phys.\ J.\  {\bf C2}, 545 (1998)
[hep-ph/9703252].
%%CITATION = HEP-PH 9703252;%%
%\href{http://www.slac.stanford.edu/spires/find/hep/www?eprint=HEP-PH/9703252}{SPIRES}


%\cite{Achasov:1999qj}
\bibitem{Achasov:1999qj}
N.~N.~Achasov and V.~V.~Gubin,
%``Final-state interaction phase difference in J/psi --> rho eta and  omega eta decays,''
Phys.\ Rev.\  {\bf D61}, 117504 (2000)
[hep-ph/9912393].
%%CITATION = HEP-PH 9912393;%%
%\href{\wwwspires?eprint=HEP-PH/9912393}{SPIRES}



%\cite{Efremov:1982sh}
\bibitem{efr82}
A.~V.~Efremov and O.~V.~Teryaev,
%``On Spin Effects In Quantum Chromodynamics,''
Sov.\ J.\ Nucl.\ Phys.\  {\bf 36}, 140 (1982)
[Yad.\ Fiz.\ {\bf 36}, 242 (1982)].
%%CITATION = SJNCA,36,140;%%
%\href{http://www.slac.stanford.edu/spires/find/hep/www?j=SJNCA%2c36%2c140}{SPIRES}

%\cite{Feldmann:2000uf}
\bibitem{fel99}
T.~Feldmann,
%``Quark structure of pseudoscalar mesons,''
Int.\ J.\ Mod.\ Phys.\  {\bf A15}, 159 (2000)
[hep-ph/9907491].
%%CITATION = HEP-PH 9907491;%%
%\href{http://www.slac.stanford.edu/spires/find/hep/www?eprint=HEP-PH/9907491}{SPIRES}




%\cite{Bisello:1991re}
\bibitem{DM2}
D.~Bisello {\it et al.}  [DM2 collaboration],
%``Study of the eta(c) decays,''
Nucl.\ Phys.\  {\bf B350}, 1 (1991).
%%CITATION = NUPHA,B350,1;%%
%\href{http://www.slac.stanford.edu/spires/find/hep/www?j=NUPHA%2cB350%2c1}{SPIRES}


%\cite{Baltrusaitis:1986mr}
\bibitem{MRK3a}
R.~M.~Baltrusaitis {\it et al.}  [Mark-III Collaboration],
%``Hadronic Decays Of The Eta(C) (2980),''
Phys.\ Rev.\  {\bf D33}, 629 (1986).
%%CITATION = PHRVA,D33,629;%%
%\href{http://www.slac.stanford.edu/spires/find/hep/www?j=PHRVA%2cD33%2c629}{SPIRES}

%\cite{Bai:1990hk}
\bibitem{MRK3}
Z.~Bai {\it et al.}  [MARK-III Collaboration],
%``Observation Of A Pseudoscalar State In J / Psi $\to$ Gamma Phi Phi Near Phi Phi Threshold,''
Phys.\ Rev.\ Lett.\  {\bf 65}, 1309 (1990).
%%CITATION = PRLTA,65,1309;%%
%\href{http://www.slac.stanford.edu/spires/find/hep/www?j=PRLTA%2c65%2c1309}{SPIRES}




%\cite{Bolz:1996sw}
\bibitem{BK96}
J.~Bolz and P.~Kroll,
%``Modelling the nucleon wave function from soft and hard processes,''
Z.\ Phys.\  {\bf A356}, 327 (1996)
[hep-ph/9603289].
%%CITATION = HEP-PH 9603289;%%
%\href{http://www.slac.stanford.edu/spires/find/hep/www?eprint=HEP-PH/9603289}{SPIRES}

%\cite{Grein:1980nw}
\bibitem{gre81}
W.~Grein and P.~Kroll,
%``Two Pion And Three Pion Cut Contributions To Nucleon-Nucleon Scattering,''
Nucl.\ Phys.\  {\bf A338}, 332 (1980).
%%CITATION = NUPHA,A338,332;%%
%\href{http://www.slac.stanford.edu/spires/find/hep/www?j=NUPHA%2cA338%2c332}{SPIRES}


%
\bibitem{sak67} J.J.\ Sakurai, Phys.\ Rev.\ Lett.\ {\bf 9}, 472 (1962); 
               Phys.\ Rev.\ Lett.\ {\bf 19}, 803 (1967).

%%CITATION = NONE;%% 


%\cite{Isgur:1975ib}
\bibitem{ruj}
N.~Isgur,
%``Why The Pseudoscalar Meson Mixing Angle Is 10-Degrees,''
Phys.\ Rev.\  {\bf D12}, 3770 (1975);
%%CITATION = PHRVA,D12,3770;%%
%\href{http://www.slac.stanford.edu/spires/find/hep/www?j=PHRVA%2cD12%2c3770}{SPIRES}
%
%\cite{Scadron:1984jw}
M.~D.~Scadron,
%``Quark Based Phenomenology Of Strong And Electromagnetic Particle Mixing,''
Phys.\ Rev.\  {\bf D29}, 2076 (1984).
%%CITATION = PHRVA,D29,2076;%%
%\href{http://www.slac.stanford.edu/spires/find/hep/www?j=PHRVA%2cD29%2c2076}{SPIRES}

\end{thebibliography}
\end{document}